\documentclass[11pt,a4paper]{article}
\pdfoutput=1
\usepackage{jcappub}
\usepackage{verbatim}
\usepackage{epsf,epsfig}
\usepackage{graphics}
\usepackage{color}
\newcommand{\beq}{\begin{equation}}
\newcommand{\eeq}{\end{equation}}
\newcommand{\bea}{\begin{eqnarray}}
\newcommand{\eea}{\end{eqnarray}}
\newcommand{\vc}[1]{{\textbf{#1}}}
\newcommand{\bmatr}{\begin{pmatrix}}
\newcommand{\ematr}{\end{pmatrix}}

\usepackage{indentfirst}
\usepackage{slashed}
\usepackage{graphics}
\usepackage{color}
\usepackage{amsmath}
\usepackage{tikz}
\usepackage{caption}
\usetikzlibrary{decorations.markings,decorations.pathmorphing,shapes.misc,snakes,patterns}
\tikzset{cross/.style={cross out, draw=black, minimum size=2*(#1-\pgflinewidth), inner sep=0pt, outer sep=0pt},
	cross/.default={1pt}}

\usepackage{rotating}
\tikzset{
	get length/.style = {
		preaction={decorate,
			decoration={ markings,
				mark = at position 1 with
				{
					\pgfkeysgetvalue{/pgf/decoration/mark info/distance from start}{\len}
					\xdef\pathlen{\len}
				}
			}
		}
	}
}

\title{Feynman Rules for Stochastic Inflationary Correlators}
\author{Marios Bounakis and}
\author{Gerasimos Rigopoulos}
\affiliation{School of Mathematics, Statistics and Physics, 
	Herschel Building, Newcastle University, 
	Newcastle upon Tyne, NE1 7RU, UK}

\abstract{We elaborate on the functional integral describing stochastic dynamics of a spectator field during inflation, comparing its diagrammatic expansion to that obtained directly from a perturbative solution of the corresponding Langevin equation. We state Feynman rules for computing arbitrary temporal $n$-point functions and perform some illustrative computations for a $\lambda\phi^4$ interaction, paying attention to the role played by a functional Jacobian determinant in the path integral. We also briefly consider the case when the field contributes to the expansion rate, making the noise multiplicative, which introduces additional vertices. }

\begin{document}
\maketitle

\section{Introduction}
After the pioneering observation by Starobinsky that the quantum behaviour of a light scalar field during inflation can be described via a local, stochastic Langevin-like equation when the field is coarse-grained over super-Hubble spatial patches \cite{Starobinsky:1986fx, Starobinsky:1994bd}, the stochastic approach to IR inflationary dynamics has been widely used by cosmologists as it offers relative conceptual and technical simplicity compared to resorting in full quantum computations. The precise relation of the stochastic approach to the underlying fundamental quantum dynamics has began to be understood \cite{Woodard:2005cw,Tsamis:2005hd, Finelli:2008zg, Garbrecht:2013coa, Garbrecht:2014dca}, while formulations that go beyond the leading IR order of de Sitter test fields' stochastic dynamics have also been achieved in \cite{Moss:2016uix, Gorbenko:2019rza, Baumgart:2019clc}, with reference \cite{Baumgart:2019clc} being the most recent study of the relation between the diagrammatics of in-in QFT and the stochastic description. 

The stochastic approach offers advantages over full quantum computations if only the IR sector is considered as it is generally simpler and, most importantly, can lead to results that go beyond perturbation theory. It can therefore be used even in situations when perturbation theory breaks down, for example in the case of a massless or very light spectator field. In such a case, even though perturbation theory does not appear to converge, an equilibrium solution exists and is easily accessible as a solution of the corresponding Fokker-Planck equation \cite{Starobinsky:1994bd}. 

Recently, the use of stochastic methods to compute IR 2-point functions for a spectator scalar in inflation has further been developed in \cite{Prokopec:2017vxx, Moreau:2019jpn, Markkanen:2019kpv, Markkanen:2020bfc, Gorbenko:2019rza}. The computation of \emph{curvature} perturbations using stochastic methods requires additional technology, the stochastic $\Delta N$ formalism \cite{Enqvist:2008kt, Fujita:2013cna, Fujita:2014tja, Vennin:2015hra, Prokopec:2019srf}, already discussed in Starobinsky's original paper \cite{Starobinsky:1986fx}.

Reference \cite{Prokopec:2017vxx} used Renormalization Group methods to compute the 2-point function of a spectator in de Sitter at two separated temporal points, while the authors of \cite{Markkanen:2019kpv, Markkanen:2020bfc} utilise the Fokker Planck equation to compute correlation functions of larger numbers of fields but again considering all of them to be at one of two temporal points. The authors of \cite{Moreau:2019jpn} use the path integral formulation of stochastic dynamics in its supersymmetric formulation to also compute unequal-time, two-point correlators of composite scalar operators and discuss their general structure. In the present work we elaborate on the path integral approach and extend the formalism developed in previous publications \cite{Garbrecht:2013coa, Garbrecht:2014dca}, stating detailed Feynman rules that allow for the straightforward perturbative computation of arbitrary $\phi$ correlators at correspondingly arbitrary temporal points. We do not employ the superfield formalism of \cite{Moreau:2019jpn} but all comparable results on 1-loop and 2-loop correlators agree with that work.      

The second aim of this paper is to compare the diagrams resulting from the path integral Feynman rules with those obtained directly from a perturbative solution of the Langevin equation.\footnote{For earlier related work on perturbative expansions for stochastic inflation see \cite{Martin:2005ir} and for a related computation within the in-in formalism see \cite{Musso:2006pt}.} They are found to be equivalent but the construction from the Langevin equation is much more laborious and less efficient. We also include a weak backreaction of the field on the expansion rate $H(\phi)$ and therefore a field dependent stochastic noise amplitude, known as multiplicative noise in the stochastic dynamics literature, briefly touching upon the extra diagrams it contributes.

The structure of the paper is as follows: In section 2 we map the Langevin equation to a path integral involving a relevant action, including the case where the expansion rate $H(\phi)$ depends weakly on the field. We discuss the role of a functional determinant appearing in the integral and demonstrate how the Fokker-Planck equation can be formally obtained directly from the path integral. In section 3 Feynman rules are obtained from the action for a quartic interaction and are used to compute the general 2- and 4-point correlators of the field to 2 loops, not necessarily at coincident times. We also compute the leading order gravitational corrections stemming from the multiplicative nature of the noise. In section 4 we perform a perturbative expansion directly from the Langevin equation and discuss the construction of the relevant diagrams which are found to be equivalent to those already obtained via the path integral. This is of course as expected but the Langevin-based diagrams are much more laborious to construct. We therefore show that the Feynman rules offer a more efficient and economic path to correlation functions. Section 5 offers a discussion and some thoughts on future research directions.     

\section{Stochastic dynamics in Inflation}

In what follows we will develop two perturbative diagrammatic expansions for $\phi$ correlators at unequal times, stating Feynman rules from the path integral (\ref{path integral 1}) in section \ref{sect:path integral} as well as the Langevin equation (\ref{Langevin}) in section 4. We show them to be equivalent but the former, utilising standard propagators and vertices, is far more economical than the later. We then apply these rules to obtain results for the two- and four-point functions at 2 and 1 loops respectively in de Sitter. We also obtain the leading order gravitational corrections to the 2-point function in the quasi de Sitter case when the dependence on $\phi$ of the expansion rate $H(\phi)$, and hence the noise amplitude, is taken into account. We first derive the Janssen-deDominicis \cite{Janssen, deDominicis} path integral formulation for the stochastic dynamics in this section.   

\subsection{Path integral formulation}
The IR dynamics of a scalar field in (quasi-)de Sitter, coarse grained over patches of physical size $\Delta r\sim 1/H$ is described by the stochastic (slow-roll) dynamical equation   
\begin{equation}\label{Langevin}
\dot{\phi} + \frac{V'}{3H} = \hbar^{1/2}\mathcal{A}[\phi(t)]\,\xi(t)
\end{equation} 
where a prime denotes a derivative w.r.t.~$\phi$, $\mathcal{A}$ satisfies 
\begin{equation}
\mathcal{A}^2 = \frac{H^3}{4\pi^2}
\end{equation}
and $\xi(t)$ is a Gaussian stochastic force term whose histories are weighted by a Gaussian probability distribution functional such that for any functional $F[\xi]$ the average over realizations of $\xi(t)$ is given by ($N$ is a normalization constant)
\begin{equation}
\langle F[\xi] \rangle = N \int \mathcal{D}\xi  \, F[\xi]\, e^{-\frac{1}{2}\int dt\, \xi(t)^2}
\end{equation}
implying 
\begin{equation}
\label{xisquared}
\langle \xi(t) \xi(t') \rangle = \, \delta(t-t')\,.
\end{equation}
for the stochastic force. In what follows we will study both cases where $H$ is constant and $\phi$ is a spectator field in de Sitter, partially developed in previous publications \cite{Garbrecht:2013coa, Garbrecht:2014dca}, as well as when it backreacts on the spacetime. Hence we will consider $H\rightarrow H(\phi)$ and assume slow roll such that 
\begin{equation}\label{Slow Roll}
H^2\simeq\frac{8\pi G}{3}V(\phi)
\end{equation}  
with de Sitter space given by $V(\phi) = V_0$. Throughout the paper we will set $c=1$ but retain $\hbar$ so as to explicitly keep track of terms related to stochastic fluctuations which are ultimately related to quantum physics during inflation.\footnote{The units of various relevant quantities are therefore $[\phi]={[{\rm mass}]^{1/2}}{[{\rm time}]^{-1/2}}\,,\quad [\mathcal{A}] = {[{\rm time}]^{-3/2}}\,,\quad [\xi] = {[{\rm time}]^{-1/2}}\,, \quad [V] = {[{\rm mass}]}{[{\rm time}]^{-3}} $
and $[\hbar]=[{\rm mass}][{\rm time}]$. }
 
To obtain expectation values of functions of the scalar $\phi(t_i)$ at times $t_i$, $\mathcal{O}[\phi(t_i)]$, on solutions $\phi_\xi$ of the stochastic equation (\ref{Langevin}) we can write       
\begin{eqnarray}
\langle \mathcal{O}[\phi(t_i)] \rangle &=&\int \!\!\mathcal{D}\xi\mathcal{D}\phi \,\, \mathcal{O}[\phi(t_i)] \, \delta\!\left[ \phi-\phi_\xi \right] e^{-\frac{1}{2}\int dt\, \xi(t)^2} \nonumber \\
&=& \int \!\!\mathcal{D}\xi\mathcal{D}\phi \,\, \mathcal{O}[\phi(t_i)] \, \delta\!\left[ \dot{\phi} + \frac{V'}{3H} -\hbar^{1/2} \mathcal{A}[\phi(t)]\,\xi(t)\right] \mathcal{J}[\phi]\,  e^{-\frac{1}{2}\int dt\, \xi(t)^2} \label{expectation 1}
\end{eqnarray}
where the resulting Jacobian determinant is 
\begin{eqnarray}
\mathcal{J}[\phi]&=&{\rm Det}\left[\frac{\delta}{\delta \phi}\left( \dot{\phi} + \frac{V'}{3H} - \hbar^{1/2}\mathcal{A}\,\xi\right)\right]\nonumber \\
&=&{\rm Det}\left[\left( \frac{d}{dt}\delta(t-t') + \left(\frac{V'}{3H}\right)' - \hbar^{1/2}\mathcal{A}'\,\xi\right)\right]
\label{Determinant}
\end{eqnarray}
where $\mathcal{A}'\equiv \partial\mathcal{A}/\partial\phi$. To obtain (\ref{expectation 1}) and (\ref{Determinant}) we made use of the functional generalization of the identity 
\begin{equation}
\delta(x-x_0) = \delta(f(x))\|f'(x)|_{x_0}\|
\end{equation} 
where $x_0$ is the solution to $f(x)=0$. The delta functional can be expressed via a functional Fourier integral as  
\begin{equation}
\delta\!\left[ \dot{\phi} + \frac{V'}{3H} - \hbar^{1/2}\mathcal{A}[\phi(t)]\,\xi(t)\right] = \int \mathcal{D}\psi \, e^{\,i\!\int \! dt \, \psi\left( \dot{\phi} + \frac{V'}{3H} - \hbar^{1/2}\mathcal{A}[\phi(t)]\,\xi(t)\right)}
\end{equation}
A convenient way to express the determinant $\mathcal{J}[\phi]$ is via the use of anti-commuting fields $\bar{c}$ and $c$
\begin{equation}
\mathcal{J}[\phi] = \int\!\mathcal{D}c\,\mathcal{D}\bar{c} \,\,e^{\int\! dt \,\bar{c}\left( \frac{d}{dt} + \left(\frac{V'}{3H}\right)' - \hbar^{1/2}\mathcal{A}'\,\xi\right)c}
\end{equation}
We can then perform the Gaussian integral over $\xi$ which leaves us with  
\begin{equation}\label{path integral 1}
\langle \mathcal{O}[\phi(t_i)] \rangle = \int\mathcal{D}\psi\mathcal{D}\phi\mathcal{D}c\mathcal{D}\bar{c} \,\, \mathcal{O}[\phi(t_i)]\,\, e^{-S}
\end{equation} 
with
\begin{equation}\label{action-canonical1}
S =\int dt\left[\frac{1}{2}\hbar\mathcal{A}^2\psi^2 -i\psi\left(\frac{d\phi}{dt}+\frac{V'}{3H} \right)+\bar{c}\left(\frac{d}{dt} +  \left(\frac{V'}{3H}\right)'-i\psi \hbar\mathcal{A}\mathcal{A}'\right)c\right]
\end{equation}
and where we have used the anti-commutativity of $\bar{c}$ and $c$ to remove one of the resulting terms containing $\bar{c}c\bar{c}c=0$. Further assuming the slow roll relation (\ref{Slow Roll}), the stochastic action becomes 
\begin{equation}\label{action-canonical2}
S = \int dt\left[\frac{1}{2}\hbar\mathcal{A}^2\psi^2 -i\psi\left(\frac{d\phi}{dt}+\frac{\left(\sqrt{V}\right)'}{\left(6\pi G\right)^{1/2}}\right)+\bar{c}\left(\frac{d}{dt} +  \frac{\left(\sqrt{V}\right)''}{\left(6\pi G\right)^{1/2}}-i\psi \hbar\mathcal{A}\mathcal{A}'\right)\!c\right].
\end{equation}
Adding source currents, 
\begin{equation}
S\rightarrow S -iJ\psi-J^q\phi-\bar{J}_c c-\bar{c}J_{\bar{c}} 
\end{equation}
where $\bar{J}_c$ and $J_{\bar{c}}$ are Grassmann valued, one obtains a generating functional $Z[J,J^q,\bar{J}_c,J_{\bar{c}}]$, which, when appropriately differentiated, provides expectation values for the fields. It should be noted that, from (\ref{expectation 1}) for $\mathcal{O}[\phi]\rightarrow 1$, 
\begin{equation}\label{Z-norm}
Z[J,0,0,0] = 1
\end{equation}
In this work we will take the potential to have the form  
\begin{equation}\label{potential}
V(\phi) = V_0 + \frac{1}{2}\frac{m^2}{\hbar^2}\phi^2 + V_{\rm int}
\end{equation}
such that
\begin{equation}
H^2 = H_0^2 + \frac{8\pi G }{6} \frac{m^2}{\hbar^2}\phi^2 +\frac{8\pi G }{3} V_{\rm int}
\end{equation}
When we discuss Feynman rules we will choose
\begin{equation}
V_{\rm int}= \frac{1}{4!} \frac{\lambda}{\hbar } \phi^4
\end{equation}
for concreteness.



\newpage

\subsection{Fokker-Planck equation}
The path integral above corresponds to a Fokker-Planck equation 
 \begin{equation}\label{Fokker-Planck}
\frac{\partial P(\phi,t)}{\partial t} = \frac{\partial}{\partial \phi}\left[\frac{V'}{3H}+ \frac{1}{4}\left(\hbar\mathcal{A}^2\right)'+\frac{1}{2}\hbar\mathcal{A}^2\frac{\partial}{\partial \phi}\right]P(\phi,t)\,.
\end{equation}    
and the correspondence is very similar to that between the Schroedinger equation and the path integral formulation of quantum mechanics \cite{Prokopec:2017vxx}. We give here a quick formal derivation as in \cite{Damgaard:1987rr}, recalling that we are assuming the Stratonovich convention $\Theta(0)=\frac{1}{2}$ in order to freely apply the normal rules of calculus. For a derivation in a general convention see \cite{Lau-Lubensky}. 

Consider an arbitrary function $\mathcal{F}(\phi)$ of the stochastic field $\phi$. Its average over different noise histories is 
\begin{equation}
\langle \mathcal{F}(\phi(t))\rangle = \int \mathcal{D}\xi \,\mathcal{F}(\phi_\xi(t))\,e^{-\frac{1}{2}\int \!dt\, \xi(t)^2} = \int\, d\phi \, \mathcal{F}(\phi) P(\phi,t)
\end{equation}  
In the first equation we wrote the expectation value as an explicit average over noise histories while in the second we utilized a time dependent probability distribution $P(\phi,t)$. Utilizing the Langevin equation we have 
\begin{equation}
\frac{d}{dt}\langle \mathcal{F}(\phi(t))\rangle = \left\langle\frac{\delta\mathcal{F}}{\delta\phi}\left(-\frac{V'}{3H}+\hbar^{1/2}\mathcal{A}\xi \right)\right\rangle 
\end{equation}  
The term on the lhs can be written in terms of the probability distribution $P(\phi,t)$ as   
\begin{equation}
\frac{d}{dt}\langle \mathcal{F}(\phi(t))\rangle =\int d\phi \, \mathcal{F}(\phi)\,\frac{\partial P}{\partial t}
\end{equation} 
The first term on the rhs is easy to deal with:
\begin{equation}
\left\langle\frac{\delta\mathcal{F}}{\delta\phi}f\right\rangle = \int d\phi \, \frac{\partial \mathcal{F}}{\partial \phi} f P = -\int d\phi \,\mathcal{F} \frac{\partial }{\partial \phi} \left(f P \right)\,.
\end{equation}  
To deal with the last term we use the functional total derivative lemma to obtain 
\begin{eqnarray}
\int \mathcal{D}\xi \, \frac{\delta}{\delta\xi}\left(\frac{\delta\mathcal{F}}{\partial \phi}\mathcal{A} \, e^{-\frac{1}{2}\int \!dt\, \xi(t)^2} \right) = 0 \Rightarrow\\
\left\langle \frac{\delta \mathcal{F} }{\delta\phi} \mathcal{A}\,\xi  \right\rangle = \int \mathcal{D}\xi \, \frac{\partial}{\partial\phi}\left(\frac{\delta\mathcal{F}}{\partial \phi}\mathcal{A}\right)\frac{\delta\phi}{\delta \xi} \, e^{-\frac{1}{2}\int \!dt\, \xi(t)^2} \Rightarrow \\
\left\langle \frac{\delta \mathcal{F} }{\delta\phi} \mathcal{A}\,\xi  \right\rangle = \left\langle \frac{\partial}{\partial\phi}\left(\frac{\delta\mathcal{F}}{\partial \phi}\mathcal{A}\right)\frac{\delta\phi}{\delta \xi}   \right\rangle = \int d\phi \, \frac{\partial}{\partial\phi}\left(\frac{\partial\mathcal{F}}{\partial \phi}\mathcal{A}\right)\frac{\delta\phi}{\delta \xi} \, P
\end{eqnarray}  
By two integrations by parts we obtain from the last term 
\begin{equation}
\left\langle \frac{\delta \mathcal{F} }{\delta\phi} \mathcal{A}\,\xi  \right\rangle =\int d\phi \,\mathcal{F}\, \frac{\partial}{\partial\phi}\left[\mathcal{A}\frac{\partial}{\partial \phi}\left(\frac{\delta\phi}{\delta \xi} P\right)\right]
\end{equation}
We still need to determine $\frac{\delta\phi(t)}{\delta\xi(t)}$. The solution to the Langevin equation can be formally written as 
\begin{equation}\label{formal-phi}
\phi(t)=\phi(t_0)+\int\limits_{t_0}^\infty\Theta(t-\tau)\left[f\left(\phi(\tau)\right)+\hbar^{1/2}\mathcal{A}(\phi(\tau))\xi(\tau)\right] 
\end{equation}
from which we obtain   
\begin{equation}
\frac{\delta \phi(t)}{\delta \xi(t)}=\hbar^{1/2}\mathcal{A}(\phi(t))\Theta(0)=\frac{1}{2}\hbar^{1/2}\mathcal{A}(\phi(t))
\end{equation}
where we utilized the Stratonovich, midpoint convention to define the ambiguous $\Theta(0)$ and all other terms in the variation of (\ref{formal-phi}) vanish as $\tau \rightarrow t$. Putting everything together and noting that $\mathcal{F}$ is arbitrary we arrive at 
\begin{equation}
\frac{\partial P(\phi,t)}{\partial t} = \frac{\partial}{\partial \phi}\left(\frac{V'}{3H}P\right) +  \frac{1}{2}\frac{\partial}{\partial \phi}\left[\hbar\mathcal{A}\frac{\partial}{\partial \phi}\left(\mathcal{A}P(\phi,t)\right)\right]\,.
\end{equation}  
which is equivalent to (\ref{Fokker-Planck}). The equilibrium solution reads
\begin{equation}
P_{\rm eq}(\phi) = \mathcal{N}H^{-3/2} \exp\left(-\frac{8\pi^2}{3\hbar}\int \frac{V'}{H^4} \, d\phi\right)
\end{equation}
where $\mathcal{N}$ is a normalization constant. 

\section{Feynman Rules from the path integral}\label{sect:path integral}
In this section we obtain Feynman rules directly from the path integral with action (\ref{action-canonical2}), also including a weak backreaction on the expansion rate $H=H(\phi)$, also corresponding to multiplicative noise in the Langevin equation.    Starting from \eqref{action-canonical2} and keeping the leading order terms in the dimensionless quantity $\chi\equiv\hbar G H_0^2/2\pi$ (which is $H_0^2/\left(2\pi M_p^2\right)$ in units where $\hbar=1$), the action in the exponent in the path integral (\ref{path integral 1}) becomes
\begin{equation}\label{action}
\begin{split}
	S=  &\frac12 \int \frac{d\omega}{2\pi} \Bigg\{ 
	\left(\begin{matrix}
		\tilde{\phi}(-\omega) \!&, \tilde{\psi}(-\omega)
	\end{matrix}\right)
	\left(\begin{matrix}
		0 &-\omega -i\slashed{m}\\
		+\omega -i\slashed{m}& \frac{\hbar H_0^3}{4\pi^2 }
	\end{matrix} \right)
	\left(\begin{matrix}
		&\tilde{\phi}(\omega)\\
		& \tilde{\psi}(\omega)
	\end{matrix}\right) \\
	&+\tilde{\bar{c}}(-\omega) \left(i \omega +\slashed{m}\right) \tilde{c}(\omega) \Bigg\} + S_{\rm int} 
\end{split}
\end{equation}
where we integrated by parts to make the kinetic term symmetric, went to Fourier space
\beq\label{Fourier}
\phi(t)=\int\limits_{-\infty}^{+\infty} \frac{d\omega}{2\pi} \, \tilde{\phi}(\omega)\,e^{i\omega t}
\eeq 
etc, and defined $\slashed{m}=\frac{m^2}{3\hbar^2 H_0}$ which has dimensions of inverse time. $S_{\rm int}$ contains the interactions
\begin{equation}
\begin{split}
\label{action-int}
S_{\rm int}= \int dt &\Bigg\{ 
-i\left[\frac{\lambda}{18\hbar\  H_0}\right]\psi \phi^3 -  \left[ \frac{\lambda}{6\hbar\  H_0} \right] \bar{c} \,\phi^2 c  \\
&+\left[\chi \ \frac{m^2}{2\hbar^2H_0}\right]\phi^2\psi^2 +\left[\chi \ \frac{\lambda}{4!\hbar  H_0}\right] \phi^4 \psi^2 + \\
&-  \left[ i \ \chi \ \frac{m^2}{\hbar^2  H_0} \right] \bar{c} \psi \phi c - \left[ i \ \chi \ \frac{\lambda}{3!\hbar H_0} \right] \bar{c} \,\psi \phi^3 c + \ldots\Bigg \} \\
\end{split}
\end{equation}
The first line in (\ref{action-int}) contains interaction terms arising in de Sitter due to self interactions of $\phi$ while the second and third lines are the leading order gravitational terms due to the field's backreaction on the spacetime geometry,  with the ellipsis denoting terms of $\mathcal{O}(\chi^2)$ or suppressed by further factors of $\frac{m^2}{H_0^2}$ and $\lambda$.

The quadratic term of (\ref{action}) involves the matrix ($\slashed{\delta}\left(\Sigma \omega_i\right)\equiv 2 \pi \, \delta\left(\Sigma \omega_i\right)$)
\begin{equation}
 \mathbf{A}_{\omega'\omega} = \left(\begin{matrix}
 0 & \omega'-i\slashed{m} \\
 -\omega'-i\slashed{m}&\frac{\hbar H_0^3}{4\pi^2}
 \end{matrix}\right)\slashed{\delta}(\omega' + \omega)
 \end{equation}
whose inverse, defined through 
\begin{equation}
\int \frac{d \omega}{2\pi}\mathbf{A}_{\omega'\omega} \cdot \mathbf{A}^{-1}_{\omega\omega''} = \mathbf{1}\times\delta(\omega'-\omega'')
\end{equation}
leads to the two point functions of the free fields 
\begin{eqnarray}
\left(\begin{matrix}
\langle\phi(\omega')\phi(\omega)\rangle & \langle\phi(\omega')\psi(\omega)\rangle \\
\langle\psi(\omega')\phi(\omega)\rangle & \langle\psi(\omega')\psi(\omega)\rangle 
\end{matrix}\right)&=&
\left(\begin{matrix}
\frac{\hbar H_0^3}{4\pi^2 (\slashed{m}^2+\omega^2)}& \frac{i}{\slashed{m}+i\omega'}\\
\frac{i}{\slashed{m}+i\omega}&0
\end{matrix}\right)\delta(\omega'+\omega)\\
&\equiv& \left(\begin{matrix}
F(\omega)& G(\omega')\\
G(\omega)&0
\end{matrix}\right)\delta(\omega'+\omega)
\end{eqnarray}
and 
\begin{equation}
\langle \tilde{\bar{c}}(\omega')\tilde{c}(\omega)\rangle = \frac{1}{ {\slashed{m}}+i\omega}\delta(\omega'+\omega)
\end{equation}
$F(\omega)$ is the Fourier transform of the free field two-point function
\begin{equation}
\langle\phi(t)\phi(t')\rangle = \int \frac{d\omega}{2\pi} F(\omega) \, e^{i\omega \left(t-t'\right)} = \frac{ \, \hbar H_0^3}{8\, \pi^2 \,\slashed{m} } \, e^{-\slashed{m} |t-t'|}
\end{equation} 
whereas the $\phi-\psi$ correlator corresponds to the retarded Green function
\begin{equation}\label{retarted}
\langle\phi(t)\psi(t')\rangle = \int \frac{d\omega}{2\pi} G(\omega) \, e^{i\omega \left(t-t'\right)} = \Theta(t-t')e^{-\slashed{m} \left(t-t'\right)}
\end{equation}
Obviously, setting $\omega \rightarrow -\omega$, or exchanging $t \leftrightarrow t'$, gives the advanced Green function. Note that the $\psi$ field always sits at an earlier time than $\phi$ in the correlators, imbuing them with a directionality, unlike the $F(t,t')$ correlator which is symmetric in $t$ and $t'$. The ghost correlator is also has a natural directionality and is simply $\mathcal{G}(\omega)=-iG(\omega)$. As we will see, it serves to maintain the normalization of the generating functional (\ref{Z-norm}). Finally, note that from (\ref{retarted}), $\Theta(0)=\frac{1}{2}$ and hence our formalism implicitly imposes the Stratonovich convention for the stochastic process.     

With the correlators/propagators described above and the interactions in (\ref{action-int}), one is lead to a diagrammatic expansion for arbitrary temporal correlators $\langle \phi(t_1)\ldots \phi(t_n)\rangle$ dictated by the following Feynman Rules:  
\begin{itemize}
	\item  Diagrams are constructed using the propagators below
	\begin{center}
		\begin{tabular}{ ccc } 
			$F(\omega)$&$\frac{H_0^3\hbar}{4\pi^2 (\slashed{m}^2+\omega^2)}$& 	
			\begin{tikzpicture}
			\filldraw[ultra thick] (1,0)circle (2pt)  -- (-1,0)circle (2pt);
			\end{tikzpicture}   \vspace{0.5cm} \\ 
			$G(\omega)$& $\frac{i}{ (\slashed{m}+i\omega)}$&
			\begin{tikzpicture}
			\filldraw [ultra thick](1,0) circle (2pt)  -- (0,0);
			\draw [snake, ultra thick   ] (0,0) -- (-1,0);
			\filldraw [ultra thick](-1,0)circle (2pt) ;
			\end{tikzpicture}  	\vspace{0.5cm}\\ 
			$\mathcal{G}(\omega)$& $\frac{1}{ ({\slashed{m}}+i\omega)}$&
			\begin{tikzpicture}
			\filldraw[ultra thick] (1,0) circle (2pt);
			\filldraw[ultra thick] (-1,0) circle (2pt);
			\draw[->, ultra thick, dashed](1,0) -- (0,0);
			\draw[ultra thick, dashed](0,0) -- (-1,0);
			\end{tikzpicture}  	\\
		\end{tabular}
	\end{center}

 Each line is associated with a frequency $\omega$ running along it. The directionality associated with $G$ is indicated by the wiggly-straight line with the two ends corresponding to the $\psi$ and $\phi$ fields respectively. If $\omega$ runs from the wiggly to the straight end it is counted as positive, whereas it is counted as $-\omega$ if it runs from the straight to the wiggly end. Alternatively, in configuration space the wiggly end corresponds to the earlier time. In ghost lines, the arrow, flowing from $c$ to $\bar{c}$, also indicates the time direction and the sence in which a frequency associated to the line is counted as positive.          
 
 \item These propagators are joined together with vertices. In the case of a spectator scalar and de Sitter spacetime, the vertices are, for a potential $V(\phi) = \lambda/{\hbar 4!} \, \phi^4$ 
 
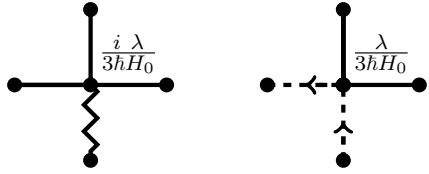
\begin{center} 
 	\begin{tikzpicture}
 \filldraw[ultra thick] (1,0)circle (2pt) -- (-1,0)circle (2pt)  -- (0,0)circle (2pt) node[align=right, above] {$\hspace{1cm} \frac{i \ \lambda}{3\hbar H_0}$} -- (0,1)circle (2pt); 
 \draw[ultra thick, snake ] (0,0) -- (0,-1);
 \filldraw[ultra thick] (0,-1)circle (2pt);
 	\end{tikzpicture}\hspace{1cm}
 	\begin{tikzpicture}
 \filldraw[ultra thick] (1,0)circle (2pt) -- (0,0)circle (2pt) -- (0,1)circle (2pt)  -- (0,0)circle (2pt) node[align=right, above] {$\hspace{1cm} \frac{ \lambda}{3\hbar H_0}$};
 \filldraw[ultra thick,dashed](0,-1) -- (0,0) ;
 \filldraw[ultra thick,dashed](-1,0) -- (0,0) ;
 \draw[->, ultra thick, dashed](0,0)--(-0.5,0);
 \draw[->, ultra thick, dashed](0,-1)--(0,-0.5); 
 \filldraw[ultra thick] (0,-1)circle (2pt);
 \filldraw[ultra thick] (-1,0)circle (2pt);
 	\end{tikzpicture} 
\end{center} 

Note that for a $\phi^n$ interaction there would be $n-1$ straight legs in the left diagram and $n-2$ straight legs in the ghost diagram with the appropriate vertex factor. There are also additional ``gravitational vertices'' stemming from the $\phi$-dependence of the noise amplitude (multiplicative noise) in \eqref{action-canonical2}, shown below: \footnote{We leave a more comprehensive analysis of $\phi$'s backreaction and other gravitational effects for future work} 

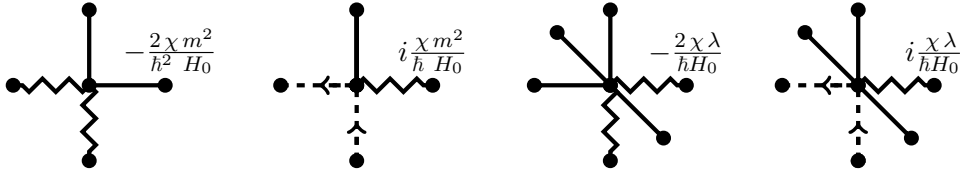
\begin{tikzpicture}
  \filldraw[ultra thick] (0,0)--(1,0)circle (2pt); 
  \filldraw[ultra thick] (0,0) circle (2pt) node[align=right, above] {$\hspace{2cm}  \ -\frac{2 \, \chi \, m^2}{\hbar^2 \ H_0}$} -- (0,1)circle (2pt) ;
  \draw[ultra thick, snake ] (0,0) -- (0,-1);
  \draw[ultra thick, snake ] (0,0) -- (-1,0);
  \filldraw[ultra thick] (0,-1)circle (2pt);
  \filldraw[ultra thick] (-1,0)circle (2pt);
\end{tikzpicture}
\begin{tikzpicture}
\filldraw[ultra thick] (0,0)circle (2pt) node[align=right, above] {\hspace{2cm}$i \frac{\, \chi \, m^2}{\hbar \ H_0}$} -- (0,1)circle (2pt) -- (0,0)circle (2pt);
\filldraw[ultra thick](0,-1)circle (2pt); 
\filldraw[ultra thick](-1,0)circle (2pt);
\filldraw[ultra thick,dashed](0,-1)-- (0,0) ;
\filldraw[ultra thick,dashed](-1,0)-- (0,0) ;
\draw[->, ultra thick, dashed](0,0)--(-0.5,0);
\draw[->, ultra thick, dashed](0,-1)--(0,-0.5);
\draw[ultra thick, snake ] (0,0) -- (1,0);
\filldraw[ultra thick] (1,0)circle (2pt);
\end{tikzpicture}  
\begin{tikzpicture}
\filldraw[ultra thick] (0,0)--(-1,0)circle (2pt); 
\filldraw[ultra thick] (0,0) circle (2pt) node[align=right, above] {$\hspace{2cm} - \frac{2\, \chi \, \lambda}{\hbar H_0}$} -- (0,1)circle (2pt);
\filldraw (0,-1)circle (2pt)  ; 
\draw[ultra thick, snake ] (0,0) -- (0,-1);
\filldraw[ultra thick](0,-1)circle (2pt);
\draw[ultra thick, snake ] (0,0) -- (1,0);
\filldraw[ultra thick] (1,0)circle (2pt);
\filldraw[ultra thick] (0,0) circle (2pt) -- (-0.7,0.7)circle (2pt);
\filldraw[ultra thick] (0,0)--(0.7,-0.7)circle (2pt); 
\end{tikzpicture}
\begin{tikzpicture}
\filldraw[ultra thick](0,0)circle (2pt) -- (0,1)circle (2pt);
\filldraw[ultra thick,dashed](0,-1)circle (2pt) -- (0,0) node[align=right, above] {$\hspace{2cm}i \frac{\, \chi \, \lambda}{\hbar H_0}$} ;
\filldraw[ultra thick,dashed](-1,0) -- (0,0) ;
\draw[->, ultra thick, dashed](0,0)--(-0.5,0);
\draw[->, ultra thick, dashed](0,-1)--(0,-0.5);
\draw[ultra thick, snake ] (0,0) -- (1,0);
\filldraw[ultra thick](1,0) circle (2pt);
\filldraw[ultra thick] (0,0) circle (2pt) -- (-0.7,0.7)circle (2pt);
\filldraw[ultra thick] (0,0)--(0.7,-0.7)circle (2pt);
\filldraw[ultra thick] (0,-1)circle (2pt);
\filldraw[ultra thick] (-1,0)circle (2pt);
\end{tikzpicture} 

\item  Frequency conservation applies at each vertex.  

\item Running inside each closed loop is a frequency $\sigma$ which is integrated over with $\int \frac{d\sigma }{2\pi}$.   

\item All external points at times $t_i$ come with a straight leg
\begin{center}
	\begin{tikzpicture}
\filldraw[ultra thick](0,0)circle (2pt) -- (1,0);
	\end{tikzpicture}
\end{center}
which attaches to a vertex on either a straight or a wiggly leg, creating the associated $F$ or $G$ propagator.   
   
\item Each external line connecting to $t_i$ carries a frequency $\omega_i$. We count it as $+\omega_i$ if it exits the diagram and $-\omega_i$ if it enters the diagram. The overall direction of frequency flow is conventional. In addition to their $F$ or $G$ factors, external lines also carry an $e^{\pm i\omega_i t_i}$ factor.   

\item Total frequency is conserved across the whole diagram and external frequencies are integrated over with $\int \frac{d\omega}{2\pi}$.  

\item Diagrams should be divided by their symmetry factor. 

\end{itemize}

We now apply these rules to perform a few illustrative computations of correlation functions by constructing the corresponding diagrams. We will focus here on the spectator field case and will utilize the extra vertices to compute some of their contributions in section \ref{Gravitational contributions}.  

\subsection{Partition function}
In the absence of external currents $J_\psi$, $\bar{J}_c$ and $J_{\bar{c}}$, the generating functional is unity by construction, see (\ref{Z-norm}). Therefore, all vacuum bubbles must vanish. Indeed, this is achieved by cancellations from the ghost loops. The partition function is expanded as:

\begin{equation}
\label{partition}
Z=1+
\begin{tikzpicture}
\filldraw [  ultra thick  ] (0,0)circle (2pt) node[align=center, below]{} ;
\draw [get length](0,0) arc (-90:90:0.5cm) ;
\draw [ ultra thick] (0,0) arc (-270:90:0.5cm);
\draw [ ultra thick] (0,1) arc (90:-90:0.5cm);
\draw[ultra thick,decoration={snake,amplitude=0.06cm,segment length={\pathlen /7}},
decorate] (0,0) arc (270:90:0.5cm);
\end{tikzpicture}  
+
\begin{tikzpicture}
\draw [ ultra thick] (0,0) arc (-270:90:0.5cm);
\filldraw [  ultra thick  ] (0,0)circle (2pt) node[align=center, below]{};
\draw [ dashed, ultra thick] (0,0.5) circle [radius=0.5];
\draw[->, ultra thick, dashed] (0,1)--(-0.01,1);
\filldraw [  ultra thick  ] (0,0)circle (2pt) node[align=center, below]{} ;
\end{tikzpicture} 
+ 
\begin{tikzpicture}
\filldraw [  ultra thick  ] (0,0)circle (2pt) node[align=center, below]{};
\draw [ dashed, ultra thick] (0,0.5) circle [radius=0.5];
\draw[->, ultra thick, dashed] (0,1)--(-0.01,1);
\draw [get length](0,0) arc (90:270:0.5cm) ;
\draw[ultra thick,decoration={snake,amplitude=0.06cm,segment length={\pathlen /7}},decorate] (0,0) arc(90:-90:0.5cm);
\draw [ ultra thick] (0,0) arc (90:270:0.5cm);
\end{tikzpicture}
+
\begin{tikzpicture}
\filldraw [  ultra thick  ] (0,0)circle (2pt) node[align=center, below]{} ;
\draw [get length](0,0) arc (90:270:0.5cm) ;
\draw[ultra thick,decoration={snake,amplitude=0.06cm,segment length={\pathlen /7}},decorate] (0,0) arc(90:-90:0.5cm);
\draw [ ultra thick] (0,-1) arc (270:90:0.5cm);
\draw [ ultra thick] (0,1) arc (90:-90:0.5cm);
\draw[ultra thick,decoration={snake,amplitude=0.06cm,segment length={\pathlen /7}},
decorate] (0,0) arc (270:90:0.5cm);
\filldraw [  ultra thick  ] (0,0)circle (2pt) node[align=center, above]{$\frac12$} ;
\end{tikzpicture}  
+
	\begin{tikzpicture}
\filldraw [  ultra thick  ] (0,0)circle (2pt) node[align=center, below]{};
\draw [ dashed, ultra thick] (0,0.5) circle [radius=0.5];
\draw[->, ultra thick, dashed] (0,1)--(-0.01,1);
\filldraw [  ultra thick  ] (0,0)circle (2pt) node[align=center, below]{} ;
\draw [get length](0,0) arc (90:270:0.5cm) ;
\draw[ultra thick,decoration={snake,amplitude=0.06cm,segment length={\pathlen /7}},decorate] (0,0) arc(90:-90:0.5cm);
\draw [ ultra thick] (0,0) arc (90:270:0.5cm);
\draw [ ultra thick] (0,0) arc (-90:360:0.8cm);
\filldraw [  ultra thick  ] (0,0)circle (2pt) node[align=center, above]{$\frac12$} ;
\end{tikzpicture}
+
\begin{tikzpicture}
\filldraw [  ultra thick  ] (0,0)circle (2pt) node[align=center, below]{} ;
\draw [get length](0,0) arc (90:270:0.5cm) ;
\draw[ultra thick,decoration={snake,amplitude=0.06cm,segment length={\pathlen /7}},decorate] (0,0) arc(90:-90:0.5cm);
\draw [ ultra thick] (0,-1) arc (270:90:0.5cm);
\draw [ ultra thick] (0,1) arc (90:-90:0.5cm);
\draw[ultra thick,decoration={snake,amplitude=0.06cm,segment length={\pathlen /7}},
decorate] (0,0) arc (270:90:0.5cm);
\draw [ ultra thick] (0,0) arc (-90:360:0.8cm);
\filldraw [  ultra thick  ] (0,0)circle (2pt) node[align=center, above]{$\frac14$} ;
\end{tikzpicture}
+ \ldots 
\end{equation}       
One should note that the bubble stemming from the $\phi^3\psi$ interaction (first bubble above), cancels the one from the $\bar{c}\phi^2c$ (second bubble above). The other four order $\lambda$ bubbles shown stem from the multiplicative  noise when $H(\phi)$ and also cancel due to ghost loops: the $\bar{c}\phi\psi c$ bubble cancels the one from $\phi^2 \psi^2$ (third and fourth bubble above) and the $\psi^2 \phi^4$ bubble cancelling the one from $\bar{c} \psi \phi^3 c$ (fifth and sixth bubble above). The symmetry factors for  $\phi^2 \psi^2$,  $\psi^2 \phi^4$ and $\bar{c} \psi \phi^3 c$ are $\frac12$, $\frac12$ and $\frac14$, respectively. This diagrammatic cancellation persists to all orders and is a consequence of the inclusion of the determinant $\mathcal{J}[\phi]$, expressed in terms of ghost fields, which ensures the correct normalization of the delta functional. More precisely, the cancellations are due to the fact that $\mathcal{G}(\omega)=-iG(\omega)$ and the corresponding factors of $i$ in the vertices.

\subsection{Two-Point Function $\langle\phi(t_1)\phi(t_2)\rangle$}

\noindent The tree-level contribution to the 2-point function $\langle\phi(t_1)\phi(t_2)\rangle$
\newline

\begin{tikzpicture}
\filldraw[ultra thick](0,0)circle (2pt) node[align=center, below] {$t_1$}--(2,0)circle (2pt) node[align=center, below] {$t_2$};
\end{tikzpicture}

\noindent is simply an F-type propagator. Applying the rules and choosing the frequency to run from right to left and defining $\sigma \equiv \frac{\omega}{\slashed{m} }$, we get

\begin{equation}
\label{twopointtree}
\langle\phi(t_1)\phi(t_2)\rangle^{(0)}=F(t_1, t_2)=\frac{\hbar \, H_0^3}{4\pi^2\, \slashed{m} }\,  \int \frac{d\sigma }{2\pi}  \, 
\frac{  e^{i\,\slashed{m} \sigma ( t_1-t_2)} }{(1+\sigma^2)}=\frac{ \, \hbar H_0^3}{8\, \pi^2 \,\slashed{m} } \, e^{-\slashed{m} |t_1-t_2|}
\end{equation}
a well known result. 
\newline

\noindent To first order in $\lambda$, in deSitter, the contributing Feynmann diagrams are: 

\vspace{0.5cm}

\noindent \textbf{The Left Seagull}\\

\noindent \begin{minipage}{3cm}
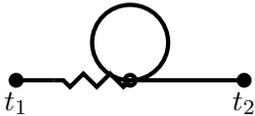
\begin{tikzpicture}
\filldraw [ultra thick](1.5,0)circle (2pt) node[align=center, below] {$t_2$} -- (0,0);
\draw [  snake,ultra thick   ] (0,0)circle (2pt) node[align=center, below]{} -- (-1,0);
\filldraw [ultra thick] (-1.5,0)circle (2pt) node[align=center, below] {$t_1$}--(-1,0);
\draw [ultra thick] (0,0.5) circle [radius=0.5];
\end{tikzpicture}
\end{minipage}
\begin{minipage}{12cm}
\begin{equation}
\begin{split}
A
&=-\frac12 \, \, \, \frac{\lambda H_0^5 \,\hbar }{24\,\, \, 4\pi^4 \, \,\slashed{m}} \int \frac{d\omega}{2\pi} \frac{e^{i \omega\left(t_1-t_2\right)}}{\left(\slashed{m}+i\omega\right)\left(\slashed{m}^2+\omega^2\right)}
\end{split}
\end{equation}
\end{minipage}\\

\noindent where we have taken frequency to run from right to left and the symmetry factor is $\frac{1}{2}$.\\

\noindent\textbf{The Right Seagull}\\

\noindent\begin{minipage}{3cm}
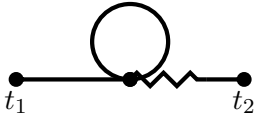
\begin{tikzpicture}
\filldraw[ultra thick] (-1.5,0)circle (2pt) node[align=center, below] {$t_1$} -- (0,0)circle (2pt);
\draw [  snake, ultra thick  ] (0,0)circle (2pt) node[align=center, below]{} -- (1,0);
\filldraw[ultra thick] (1,0)--(1.5,0)circle (2pt) node[align=center, below] {$t_2$};
\draw [ ultra thick] (0,0.5) circle [radius=0.5];
\end{tikzpicture}
\end{minipage}
\begin{minipage}{12cm}
\begin{equation}
\begin{split}
B
&=-\frac12 \, \, \, \frac{\lambda H_0^5\,\hbar }{24\,\, \, 4\pi^4 \, \,\slashed{m}} \int \frac{d\omega}{2\pi} \frac{e^{i \omega\left(t_1-t_2\right)}}{\left(\slashed{m}^2+\omega^2\right)\left(\slashed{m}-i\omega\right)}
\end{split}
\end{equation}
\end{minipage}\\

\noindent where again frequency was taken to run from right to left. Together they give 

\begin{equation}
A+B=- \frac{\lambda H_0^5 }{24 \pi^4} \ \frac{\hbar}{4} \int \frac{d\omega}{2\pi} \, \frac{e^{i\omega\left(t_1-t_2\right)}}{\left(\slashed{m}^2 +\omega^2\right)^2}
\end{equation}
which is symmetric in $t_1\leftrightarrow t_2$.

Two more diagrams can be formed at order $\lambda$ with the existing vertices, one including a closed $G$ propagator line

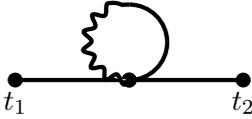
\begin{minipage}{3cm}
\begin{tikzpicture}
\filldraw[ultra thick] (-1.5,0)circle (2pt) node[align=center, below] {$t_1$} -- (0,0)circle (2pt);
\filldraw [  ultra thick  ] (0,0)circle (2pt) node[align=center, below]{} -- (1.5,0)circle (2pt) node[align=center, below] {$t_2$};
\draw [ ultra thick] (0,1) arc (90:-90:0.5cm);
\draw [get length,white](0,0) arc (270:90:0.5cm) ;
\draw[ultra thick,decoration={snake,amplitude=0.06cm,segment length={\pathlen /7}},
decorate] (0,0) arc (270:90:0.5cm);
\end{tikzpicture}
\end{minipage}
\begin{minipage}{12cm}
\begin{equation}
\begin{split}
C
&= \frac{-\lambda H_0^5}{24\, \pi^4} \frac{\hbar}{2} \int \frac{d\omega}{2\pi} \,\, \frac{e^{i\omega \left(t_1-t_2\right)}}{\left(\slashed{m}^2+\omega^2\right)^2} \Theta(0)
\end{split}
\end{equation}
\end{minipage}

\noindent and the ghost-loop diagram  

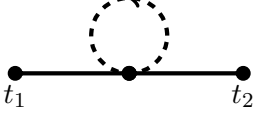
\begin{minipage}{3cm}
\begin{tikzpicture}
\filldraw[ultra thick] (-1.5,0)circle (2pt) node[align=center, below] {$t_1$} -- (0,0)circle (2pt);
\filldraw [  ultra thick  ] (0,0)circle (2pt) node[align=center, below]{} -- (1.5,0)circle (2pt) node[align=center, below] {$t_2$};
\draw [ dashed, ultra thick] (0,0.5) circle [radius=0.5];
\draw[->, ultra thick, dashed] (0,1)--(-0.01,1);
\end{tikzpicture}
\end{minipage}
\begin{minipage}{12cm}
\begin{equation}
\begin{split}
\label{ghostloop}
 G
&=\frac{\lambda H_0^5}{24\, \pi^4\,} \frac{\hbar}{2} \int \frac{d\omega}{2\pi} \,\, \frac{e^{i\omega \left(t_1-t_2\right)}}{\left(\slashed{m}^2+\omega^2\right)^2} \Theta(0)
\end{split}
\end{equation}
\end{minipage}

\noindent where $\Theta(0)=\frac{1}{2}$ with the ghost loop acting to precisely cancel the retarded propagator loop, as expected.\footnote{This is true for any assignment of a value for $\Theta(0)$, not only for $\Theta(0)=1/2$, as our formalism implies here, reflecting the fact that for additive noise, $H=H_0$, results are independent of the stochastic discretization prescription: Stratonovich, Ito or otherwise.} Then, the first order in $\lambda$ contribution to the two point function is given by: 

\begin{equation}
\begin{split}
\label{propagatorl1}
\langle\phi(t_1)\phi(t_2)\rangle^{(1)}
&=\frac{\hbar \, H_0^3}{4\pi^2 \,  \slashed{m}} \left[-\frac{\lambda \ H_0^2 }{4\pi^2\, 6\slashed{m}^2 }  \right]
\int \frac{d\sigma}{2\pi} \, \frac{e^{i \slashed{m} \sigma\left(t_1-t_2\right)}}{\left(1 +\sigma^2\right)^2} \\
&=\frac{\hbar \, H_0^3}{4\pi^2 \,  \slashed{m}} \left[-\frac{\lambda \ H_0^2 }{4\pi^2\, 6\slashed{m}^2 }  \right] \frac14  \left(1+\slashed{m}|\Delta t|\right) e^{-\slashed{m}|\Delta t|}
 \end{split}
\end{equation}
where $\Delta t = t_1-t_2$.

\noindent {\bf NNLO deSitter contributions}
\\
\noindent To second order in $\lambda$, there are 3 distinct topologies of connected Feynman diagrams, contributing with two loops: 

\noindent \textbf{The Symmetric Sunset}

Taking frequency $\omega$ to run trough the diagram from right to left and noting that the diagram's symmetry factor is $\frac{1}{6}$, we have  

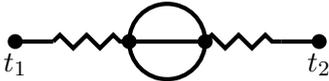
\begin{minipage}{5cm}
	\begin{tikzpicture}
	\filldraw[ultra thick](4.5,0)circle (2pt) node[align=center, below] {$t_2$} ;
	\filldraw[ultra thick](3,0)circle (2pt) node[align=center, below] {$\hspace{0.3cm}  $} ;
	\filldraw[ultra thick](2,0)circle (2pt) node[align=center, below] {$\hspace{-0.3cm} $} ;
	\filldraw[ultra thick](0.5,0)circle (2pt) node[align=center, below] {$t_1$} ;
	\filldraw [ultra thick](2,0) -- (3,0);
	\filldraw [ultra thick](0.5,0) -- (1,0);
	\filldraw [ultra thick](4,0) -- (4.5,0);
	\draw [  snake,ultra thick   ] (1,0)-- (2,0);
	\draw [  snake,ultra thick   ] (3,0)-- (4,0);
	\draw [ultra thick] (2.5,0) circle [radius=0.5];
	\end{tikzpicture}
\end{minipage}
\begin{minipage}{10cm}
	\begin{equation}
	\begin{split}
	\label{symsun}
	\mathcal{SS}&=\frac16 \left(i\frac{\lambda}{3H_0 \, \hbar}\right)^2 \int\frac{d\omega}{2\pi}\frac{e^{i \omega\left(t_1-t_2\right)}\,\,i^2}  {\left(\slashed{m}-i\omega\right)\left(\slashed{m}+i\omega \right) }\mathcal{I}_{\mathcal{SS}}(\omega)
	\\
	&=\frac16 \frac{ \lambda^2 H_0^7 \,6 \hbar }{9\slashed{m}^2 \,\, \, \left(8\pi^2\right)^3 } \int \frac{d\omega}{2\pi} \frac{ e^{i \omega\left(t_1-t_2\right)} ( \slashed{m}^2+\omega^2 ) }{ \left(\slashed{m}^2+\omega^2\right)^2 \left[ \left(3\slashed{m}\right)^2 +\omega^2\right] }
	\end{split}
	\end{equation}
\end{minipage}

\noindent where the $\mathcal{I}_{\mathcal{SS}}(\omega)$ factor involves integrations from the two loops
\begin{equation}
\mathcal{I}_{\mathcal{SS}}(\omega)=\left(\frac{H_0^3\hbar}{4\pi^2}\right)^3\int \frac{d\sigma}{2\pi}\frac{d\rho}{2\pi} \frac{1}{\left(\slashed{m}^2+\sigma^2\right)\left(\slashed{m}^2+\rho^2\right)\left(\slashed{m}^2+\left(\sigma+\rho+\omega\right)^2\right)}
\end{equation} 
\noindent $\sigma$ runs clockwise in the upper loop and $\rho$ runs anti-clockwise in the lower loop. 

\noindent \textbf{The Left Sunset}

\noindent With frequency running from right to left again and a symmetry factor of $\frac{1}{2}$, we have 

\begin{minipage}{5cm}
	\begin{tikzpicture}
	\filldraw[ultra thick](4.5,0)circle (2pt) node[align=center, below] {$t_2$} ;
	\filldraw[ultra thick](3,0)circle (2pt) node[align=center, below] {$\hspace{0.3cm}$} ;
	\filldraw[ultra thick](2,0)circle (2pt) node[align=center, below] {$\hspace{-0.3cm}$} ;
	\filldraw[ultra thick](0.5,0)circle (2pt) node[align=center, below] {$t_1$} ;
	\filldraw [ultra thick](2,0) -- (2.5,0);
	\filldraw [ultra thick](0.5,0) -- (1,0);
	\filldraw [ultra thick](3,0) -- (4.5,0);
	\draw [  snake,ultra thick   ] (1,0)-- (2,0);
	\draw [  snake,ultra thick   ] (2.5,0)-- (3,0);
	\draw [ultra thick] (2.5,0) circle [radius=0.5];
	\end{tikzpicture}
\end{minipage}
\begin{minipage}{10cm}
	\begin{equation}
	\begin{split}
	\label{leftsun}
	\mathcal{LS}&=\frac12 \left(i\frac{\lambda}{3H_0  \hbar}\right)^2\left(\frac{H_0^3\hbar}{4\pi^2}\right)\int \frac{d\omega}{2\pi} \frac{e^{i \omega\left(t_1-t_2\right)}\,\,i}{\left(\slashed{m}^2+\omega^2\right)\left(\slashed{m}+i\omega\right)} \mathcal{I}_{\mathcal{LS}}(\omega) \\
	&=\frac{ \lambda^2 H_0^7 \,\hbar }{9\slashed{m}^2  \left(8\pi^2\right)^3 } \int \frac{d\omega}{2\pi} \frac{e^{i \omega\left(t_2-t_1\right)}  \ (\slashed{m}-i\omega)(3\slashed{m}-i\omega)  }{\left(\slashed{m}^2+\omega^2\right)^2 \left[ \left(3\slashed{m}\right)^2 +\omega^2\right] }
	\end{split}
	\end{equation}
\end{minipage}

\noindent where the loop integral now is
\begin{equation}
\mathcal{I}_{\mathcal{LS}}(\omega) = \left(\frac{H_0^3\hbar}{4\pi^2}\right)^2\int\frac{d\rho}{2\pi}\frac{d\sigma}{2\pi}\frac{i}{\left(\slashed{m}^2+\sigma^2\right)\left(\slashed{m}^2+\rho^2\right)\left(\slashed{m}+i\left(\sigma+\rho+\omega\right)\right)}
\end{equation}
with $\sigma$ and $\rho$ running in the loops as above.

\noindent \textbf{The Right Sunset}\\
Similarly 

\begin{minipage}{5cm}
	\begin{tikzpicture}
	\filldraw[ultra thick](4.5,0)circle (2pt) node[align=center, below] {$t_2$} ;
	\filldraw[ultra thick](3,0)circle (2pt) node[align=center, below] {$\hspace{0.3cm}$} ;
	\filldraw[ultra thick](2,0)circle (2pt) node[align=center, below] {$\hspace{-0.3cm}$} ;
	\filldraw[ultra thick](0.5,0)circle (2pt) node[align=center, below] {$t_1$} ;
	\filldraw [ultra thick](4,0) -- (4.5,0);
	\filldraw [ultra thick](2.7,0) -- (3,0);
	\filldraw [ultra thick](0.5,0) -- (2,0);
	\draw [  snake,ultra thick   ] (2,0)-- (2.7,0);
	\draw [  snake,ultra thick   ] (3,0)-- (4,0);
	\draw [ultra thick] (2.5,0) circle [radius=0.5];
	\end{tikzpicture}
\end{minipage}
\begin{minipage}{10cm}
	\begin{equation}
	\begin{split}
	\label{rightsun}
	\mathcal{RS}&=\frac12 \left(i\frac{\lambda}{3H_0 \hbar}\right)^2\int \frac{d\omega}{2\pi} \frac{e^{i \omega\left(t_1-t_2\right)}\,\,i}{\left(\slashed{m}^2+\omega^2\right)\left(\slashed{m}-i\omega\right)} \mathcal{I}_{\mathcal{RS}}(\omega) \\
	&=\frac{ \lambda^2 H_0^7 \hbar }{9\slashed{m}^2 \,\, \, \left(8\pi^2\right)^3 } \int \frac{d\omega}{2\pi} \frac{e^{i \omega\left(t_2-t_1\right)}  \ (\slashed{m}+i\omega)(3\slashed{m}+i\omega)  }{\left(\slashed{m}^2+\omega^2\right)^2 \left[ \left(3\slashed{m}\right)^2 +\omega^2\right] }
	\end{split}
	\end{equation}
\end{minipage}\\
and
\begin{equation}
\mathcal{I}_{\mathcal{RS}}(\omega) =\left(\frac{H_0^3\hbar}{4\pi^2}\right) \int\frac{d\rho}{2\pi}\frac{d\sigma}{2\pi}\frac{i}{\left(\slashed{m}^2+\sigma^2\right)\left(\slashed{m}^2+\rho^2\right)\left(\slashed{m}-i\left(\sigma+\rho+\omega\right)\right)}.
\end{equation}
Notice the minus signs in the frequencies that enter the propagators, resulting from their flow from the straight to the wiggly ends of the lines.

\noindent \textbf{The Symmetric Double Seaguls}\\
\begin{minipage}{5cm}
	\begin{tikzpicture}
	\filldraw[ultra thick](5,0)circle (2pt) node[align=center, below] {$t_2$} ;
	\filldraw[ultra thick](3.5,0)circle (2pt) node[align=center, below] {$$} ;
	\filldraw[ultra thick](2,0)circle (2pt) node[align=center, below] {$$} ;
	\filldraw[ultra thick](0.5,0)circle (2pt) node[align=center, below] {$t_1$} ;
	\filldraw [ultra thick](2,0) -- (3.5,0);
	\filldraw [ultra thick](0.5,0) -- (1,0);
	\filldraw [ultra thick](4.5,0) -- (5,0);
	\draw [  snake,ultra thick   ] (1,0)-- (2,0);
	\draw [  snake,ultra thick   ] (3.5,0)-- (4.5,0);
	\draw [ultra thick] (2,0.5) circle [radius=0.5];
	\draw [ultra thick] (3.5,0.5) circle [radius=0.5];
	\end{tikzpicture}
\end{minipage}
\begin{minipage}{10cm}
	\begin{equation}
	\begin{split}
	\label{symsseagull}
	\mathcal{SDS}&=\frac14 \left(i\frac{\lambda}{3H \hbar}\right)^2\left(\frac{1}{2}\frac{H_0^3\hbar}{4\pi^2}\right)^2 \int \frac{d\omega}{2\pi}\frac{e^{i \omega\left(t_1-t_2\right)}\,\, \frac{H_0^3\hbar}{4\pi^2} i^2}{\left(\slashed{m}+i\omega\right)\left(\slashed{m}^2+\omega^2\right)\left(\slashed{m}-i\omega\right)}\\
	&=\frac14 \, \, \, \frac{2 \, \lambda^2 H_0^7 \,\hbar }{9\slashed{m}^2 \,\, \, \left(8\pi^2\right)^3 } \int \frac{d\omega}{2\pi} \frac{e^{i \omega\left(t_2-t_1\right)}  }{\left(\slashed{m}^2+\omega^2\right)^2 }
	\end{split}
	\end{equation}
\end{minipage}

\noindent \textbf{The Left Double Seagulls}\\
\begin{minipage}{5cm}
	\begin{tikzpicture}
	\filldraw[ultra thick](5,0)circle (2pt) node[align=center, below] {$t_2$} ;
	\filldraw[ultra thick](3.5,0)circle (2pt) node[align=center, below] {$$} ;
	\filldraw[ultra thick](2,0)circle (2pt) node[align=center, below] {$$} ;
	\filldraw[ultra thick](0.5,0)circle (2pt) node[align=center, below] {$t_1$} ;
	\filldraw [ultra thick](2,0) -- (2.7,0);
	\filldraw [ultra thick](0.5,0) -- (1.2,0);
	\filldraw [ultra thick](3.5,0) -- (5,0);
	\draw [  snake,ultra thick   ] (1.2,0)-- (2,0);
	\draw [  snake,ultra thick   ] (2.7,0)-- (3.5,0);
	\draw [ultra thick] (2,0.5) circle [radius=0.5];
	\draw [ultra thick] (3.5,0.5) circle [radius=0.5];
	\end{tikzpicture}
\end{minipage}
\begin{minipage}{10cm}
	\begin{equation}
	\begin{split}
	\label{leftsseagull}
	\mathcal{LDS}&=\frac14 \left(i\frac{\lambda}{3H \hbar}\right)^2 \left(\frac12 \frac{H_0^3\hbar}{4\pi^2}\right)^2\int \frac{\omega}{2\pi} \frac{e^{i \omega\left(t_1-t_2\right)}\,\, \frac{H_0^3\hbar}{4\pi^2} i^2}{\left(\slashed{m}+i\omega\right)\left(\slashed{m}+i\omega\right)\left(\slashed{m}^2+\omega^2\right)}\\
	&=\frac14 \, \, \, \frac{2 \, \lambda^2 H_0^7 \,\hbar }{9\slashed{m}^2 \,\, \, \left(8\pi^2\right)^3 } \int \frac{d\omega}{2\pi} \frac{e^{i \omega\left(t_2-t_1\right)}  }{\left(\slashed{m}^2+\omega^2\right) \left(\slashed{m}+i\omega\right)^2}
	\end{split}
	\end{equation}
\end{minipage}

\noindent \textbf{The Right Double Seaguls}\\
\begin{minipage}{5cm}
	\begin{tikzpicture}
	\filldraw[ultra thick](5,0)circle (2pt) node[align=center, below] {$t_2$} ;
	\filldraw[ultra thick](3.5,0)circle (2pt) node[align=center, below] {$$} ;
	\filldraw[ultra thick](2,0)circle (2pt) node[align=center, below] {$$} ;
	\filldraw[ultra thick](0.5,0)circle (2pt) node[align=center, below] {$t_1$} ;
	\filldraw [ultra thick](2.8,0) -- (3.5,0);
	\filldraw [ultra thick](0.5,0) -- (2,0);
	\filldraw [ultra thick](4.5,0) -- (5,0);
	\draw [  snake,ultra thick   ] (2,0)-- (2.8,0);
	\draw [  snake,ultra thick   ] (3.5,0)-- (4.5,0);
	\draw [ultra thick] (2,0.5) circle [radius=0.5];
	\draw [ultra thick] (3.5,0.5) circle [radius=0.5];
	\end{tikzpicture}
\end{minipage}
\begin{minipage}{10cm}
	\begin{equation}
	\begin{split}
	\label{rightsseagull}
	\mathcal{RDS}&=\frac14 \left(i\frac{\lambda}{3H \hbar}\right)^2 \left(\frac12 \frac{H_0^3\hbar}{4\pi^2}\right)^2\int \frac{\omega}{2\pi} \frac{e^{i \omega\left(t_1-t_2\right)}\,\, \frac{H_0^3\hbar}{4\pi^2} i^2}{\left(\slashed{m}^2+\omega^2\right)\left(\slashed{m}-i\omega\right)\left(\slashed{m}-i\omega\right)}\\
	&=\frac14 \, \, \, \frac{2 \, \lambda^2 H_0^7 \,\hbar }{9\slashed{m}^2 \,\, \, \left(8\pi^2\right)^3 } \int \frac{d\omega}{2\pi} \frac{e^{i \omega\left(t_1-t_2\right)}  }{\left(\slashed{m}^2+\omega^2\right) \left(\slashed{m}-i\omega\right)^2}
	\end{split}
	\end{equation}
\end{minipage}

\noindent \textbf{The Left Cactus}\\

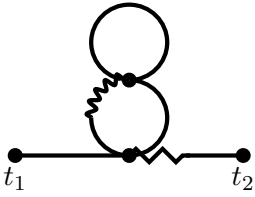
\begin{minipage}{5cm}
	\begin{tikzpicture}
	\filldraw[ultra thick](-1.5,0)circle(2pt) node[align=center, below] {$t_1$};
	\filldraw[ultra thick](0,0)circle (2pt) node[align=center, below] {$$};
	\filldraw[ultra thick](1.5,0)circle (2pt) node[align=center, below] {$t_2$};
	\filldraw[ultra thick](0,1)circle (2pt) node[align=center, below] {$$};
	\filldraw[ultra thick] (-1.5,0)-- (0,0);
	\draw [  snake,ultra thick   ] (0,0)-- (0.8,0);
	\filldraw[ultra thick] (0.75,0)-- (1.5,0);
	\draw [ultra thick] (0,1.5) circle [radius=0.5];
	\draw [ ultra thick] (-0.5,0.5) arc (-180:90:0.5cm);
	\draw [get length,white](-0.5,0.5) arc (180:90:0.5cm) ;
	\draw[ultra thick,decoration={snake,amplitude=0.06cm,segment length={\pathlen /5}},
	decorate] (-0.5,0.5) arc (180:90:0.5cm);
	\end{tikzpicture}
	\end{minipage}
\begin{minipage}{12cm}
	\begin{equation}
	\begin{split}
	\label{leftcactus}
	\mathcal{LC}&=\frac12 \left(i\frac{\lambda}{3H_0 \hbar}\right)^2 \mathcal{I}_{\mathcal{LC}}\int \frac{d\omega}{2\pi} \frac{e^{i \omega\left(t_1-t_2\right)} \,\, \frac{H_0^3\hbar}{4\pi^2}\,\, i}{\left(\slashed{m}^2+\omega^2\right)\left(\slashed{m}-i\omega\right)}   \\
	&=\frac12 \, \, \, \frac{\lambda^2 H_0^7 \,\hbar }{9\slashed{m}^3 \,\, \, \left(8\pi^2\right)^3 } \int \frac{d\omega}{2\pi} \frac{e^{i \omega\left(t_2-t_1\right)} \, \, (\slashed{m}+i\omega)}{\left(\slashed{m}^2+\omega^2\right)^2} 
	\end{split}
	\end{equation}
\end{minipage}
\\
where 
\begin{equation}
\mathcal{I}_{\mathcal{LC}}=\frac12 \frac{H_0^3\hbar}{4\pi^2}\int\frac{d\sigma}{2\pi}\frac{i\,\frac{H_0^3\hbar}{4\pi^2}}{\left(\slashed{m}+i\sigma\right)\left(\slashed{m}^2+\sigma^2\right)}
\end{equation}
is the $\omega$-independent constant stemming from the loop integrations.

\noindent \textbf{The Right Cactus}\\

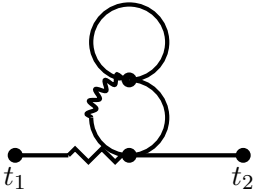
\begin{minipage}{5cm}
	\begin{tikzpicture}
	\filldraw[ultra thick](-1.5,0)circle(2pt) node[align=center, below] {$t_1$};
	\filldraw[ultra thick](0,0)circle (2pt) node[align=center, below] {$$};
	\filldraw[ultra thick](1.5,0)circle (2pt) node[align=center, below] {$t_2$};
	\filldraw[ultra thick](0,1)circle (2pt) node[align=center, below] {$$};
	\filldraw[ultra thick] (-1.5,0)-- (-0.8,0);
	\draw [  snake,ultra thick   ] (-0.8,0)-- (0,0);
	\filldraw[ultra thick] (0,0)-- (1.5,0);
	\draw [ultra thick] (0,1.5) circle [radius=0.5];
	\draw [ ultra thick] (-0.5,0.5) arc (-180:90:0.5cm);
	\draw [get length,white](-0.5,0.5) arc (180:90:0.5cm) ;
	\draw[ultra thick,decoration={snake,amplitude=0.06cm,segment length={\pathlen /5}},
	decorate] (-0.5,0.5) arc (180:90:0.5cm);
	\end{tikzpicture}
\end{minipage}
\begin{minipage}{12cm}
	\begin{equation}
		\begin{split}
	\label{rightcactus}
	\mathcal{RC}&=\frac12 \left(i\frac{\lambda}{3H_0 \hbar}\right)^2 \mathcal{I}_{\mathcal{RC}}\int \frac{d\omega}{2\pi} \frac{e^{i \omega\left(t_1-t_2\right)} \,\, i  \,\, \frac{H_0^3\hbar}{4\pi^2}}{\left(\slashed{m}+i\omega\right)\left(\slashed{m}^2+\omega^2\right)}   \\
	&=\frac12 \, \, \, \frac{\lambda^2 H_0^7 \,\hbar }{9\slashed{m}^3 \,\, \, \left(8\pi^2\right)^3 } \int \frac{d\omega}{2\pi} \frac{e^{i \omega\left(t_1-t_2\right)} \, \, (\slashed{m}-i\omega)}{\left(\slashed{m}^2+\omega^2\right)^2} 
	\end{split}
	\end{equation}
\end{minipage}\\
with $\mathcal{I}_{\mathcal{RC}}=\mathcal{I}_{\mathcal{LC}}$. 

Cacti involving a closed $G$ loop at their top are cancelled by the corresponding closed ghost loops. Therefore, adding up the above connected diagrams, we get for the 2-point function
\begin{equation}\label{propagatorl2}
\begin{split}
 \langle\phi(t) \phi(t')\rangle^{(2)} =& \frac{\hbar \, H_0^3}{4\pi^2 \,  \slashed{m}} \left[-\frac{\lambda \ H_0^2 }{4\pi^2\, 6\slashed{m}^2 }  \right]^2 \int \frac{d\sigma}{2\pi} \frac{e^{i\slashed{m} \sigma\left(t_1-t_2\right)} }{\left(1+\sigma^2\right)^2 } \left[ -\frac{1}{4} +\frac{1}{1+\sigma^2  } +  \frac{8}{9+\sigma^2 }   \right] \\
=&\frac{\hbar \, H_0^3}{4\pi^2 \,  \slashed{m}} \left[-\frac{\lambda \ H_0^2 }{4\pi^2\, 6\slashed{m}^2 }  \right]^2\frac{1}{48}\left[3\left(2 + \slashed{m}|\Delta t|\right)\left(1+ \slashed{m}|\Delta t|\right)e^{-\slashed{m}|\Delta t|}+e^{-3\slashed{m}|\Delta t|}\right]
\end{split}
\end{equation}
It is clear from these results that for temporal correlators the relevant expansion parameter is $\frac{\lambda \ H_0^2 }{4\pi^2\, 6\slashed{m}^2}$ as noted in \cite{Burgess:2010dd, Garbrecht:2013coa}.

It is worth noting that, by utilising de Sitter invariance, the temporal correlator can provide the spatial 2-point function $\langle \phi(\vc{r}_1)\phi(\vc{r}_2)\rangle$, a quantity of more direct observational interest, by replacing \cite{Starobinsky:1994bd, Garbrecht:2013coa, Markkanen:2019kpv, Markkanen:2020bfc}
\begin{equation}
\slashed{m}\Delta t \rightarrow 2\frac{\slashed{m}}{H}\ln \left(aH|\vc{r}_1-\vc{r}_2|\right) 
\end{equation}   
in the expressions (\ref{propagatorl1}) and (\ref{propagatorl2}), where $t$ is some arbitrary time of interest. This applies when equilibrium has been reached, an assumption implicit in all our computations in this work.       

\subsection{Four-Point Function} 
\label{picandies}
The Feynman rules can easily be applied to compute any higher point function at arbitrary times. For illustration we compute here the one-loop connected 4-point function, given by ``candy'' diagrams. The existence of $F$ and $G$ lines introduces different topologies that contribute to the final result. 

\subsubsection{$F$ Candies}
For the first set of diagrams we choose to connect $t_1$ to $t_3$ without having to go through the loop. There are four such diagrams, differing in the distribution of external $F$ and $G$ lines, shown below. For each diagram the symmetry factor is $2$ due to the internal $F$ propagators.

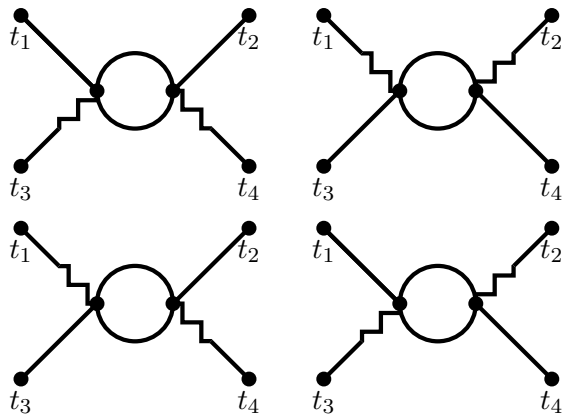
\begin{table}[h]
	\begin{center}
	\begin{tabular}{cc} 

		\begin{tikzpicture}
		\filldraw[ultra thick] (0,0)circle (2pt) node[align=center, right]{$$};
		\filldraw[ultra thick] (1,0)circle (2pt) node[align=center, left]{$$}; 
		\filldraw[ultra thick] (-1,1)circle (2pt) node[align=center, below]{$t_1$};
		\filldraw[ultra thick] (-1,-1)circle (2pt) node[align=center, below]{$t_3$};
		\filldraw[ultra thick] (2,1)circle (2pt) node[align=center, below]{$t_2$};
		\filldraw[ultra thick] (2,-1)circle (2pt) node[align=center, below]{$t_4$};
		\filldraw[ultra thick](-1,1)--(0,0);
		\filldraw[ultra thick](-1,-1)--(-0.5,-0.5);
		\filldraw[ultra thick](2,1)--(1,0);
		\filldraw[ultra thick](2,-1)--(1.5,-0.5);
		\filldraw[ultra thick](-1,1)--(0,0);
		\draw [ ultra thick] (0.5,0) circle(0.5);
		\draw[ultra thick, snake ] (0,0) -- (-0.5,-0.5);
		\draw[ultra thick, snake ] (1,0) -- (1.5,-0.5);
		\end{tikzpicture}	
		&
		\begin{tikzpicture}
		\filldraw[ultra thick] (0,0)circle (2pt) node[align=center, right]{$$};
		\filldraw[ultra thick] (1,0)circle (2pt) node[align=center, left]{$$}; 
		\filldraw[ultra thick] (-1,1)circle (2pt) node[align=center, below]{$t_1$};
		\filldraw[ultra thick] (-1,-1)circle (2pt) node[align=center, below]{$t_3$};
		\filldraw[ultra thick] (2,1)circle (2pt) node[align=center, below]{$t_2$};
		\filldraw[ultra thick] (2,-1)circle (2pt) node[align=center, below]{$t_4$};
		\filldraw[ultra thick](1,0)--(2,-1);
		\filldraw[ultra thick](-1,-1)--(0,0);
		\filldraw[ultra thick](2,1)--(1.5,0.5);
		\filldraw[ultra thick](-1,1)--(-0.5,0.5);
		\draw [ ultra thick] (0.5,0) circle(0.5);
		\draw[ultra thick, snake ] (0,0) -- (-0.5,0.5);
		\draw[ultra thick, snake ] (1,0) -- (1.5,0.5);
		\end{tikzpicture}
		\\

		\begin{tikzpicture}
		\filldraw[ultra thick] (0,0)circle (2pt) node[align=center, right]{$$};
		\filldraw[ultra thick] (1,0)circle (2pt) node[align=center, left]{$$}; 
		\filldraw[ultra thick] (-1,1)circle (2pt) node[align=center, below]{$t_1$};
		\filldraw[ultra thick] (-1,-1)circle (2pt) node[align=center, below]{$t_3$};
		\filldraw[ultra thick] (2,1)circle (2pt) node[align=center, below]{$t_2$};
		\filldraw[ultra thick] (2,-1)circle (2pt) node[align=center, below]{$t_4$};
		\filldraw[ultra thick](-1,1)--(-0.5,0.5);
		\filldraw[ultra thick](-1,-1)--(0,0);
		\filldraw[ultra thick](2,1)--(1,0);
		\filldraw[ultra thick](1.5,-0.5)--(2,-1);
		\draw [ ultra thick] (.5,0) circle(0.5);
		\draw[ultra thick, snake ] (0,0) -- (-0.5,0.5);
		\draw[ultra thick, snake ] (1,0) -- (1.5,-0.5);
		\end{tikzpicture}			
		&
		\begin{tikzpicture}
		\filldraw[ultra thick] (0,0)circle (2pt) node[align=center, right]{$$};
		\filldraw[ultra thick] (1,0)circle (2pt) node[align=center, left]{$$}; 
		\filldraw[ultra thick] (-1,1)circle (2pt) node[align=center, below]{$t_1$};
		\filldraw[ultra thick] (-1,-1)circle (2pt) node[align=center, below]{$t_3$};
		\filldraw[ultra thick] (2,1)circle (2pt) node[align=center, below]{$t_2$};
		\filldraw[ultra thick] (2,-1)circle (2pt) node[align=center, below]{$t_4$};
		\filldraw[ultra thick](-1,1)--(0,0);
		\filldraw[ultra thick](-1,-1)--(-0.5,-0.5);
		\filldraw[ultra thick](2,1)--(1.5,0.5);
		\filldraw[ultra thick](2,-1)--(1,0);
		\filldraw[ultra thick](-1,1)--(0,0);
		\draw [ ultra thick] (0.5,0) circle(0.5);
		\draw[ultra thick, snake ] (0,0) -- (-0.5,-0.5);
		\draw[ultra thick, snake ] (1,0) -- (1.5,0.5);
		\end{tikzpicture}
		\\

	\end{tabular}
\end{center}

\end{table}

Focusing on the upper left diagram, we assign incoming frequencies $\omega_1$ and $\omega_2$ to the $F$ lines attached to $t_1$ and $t_2$ and outgoing frequencies $\omega_3$ and $\omega_4$ to the $G$ lines connected to $t_3$ and $t_4$, while we assign frequency $\sigma$ to run counter-clockwise in the loop. Applying the rules we have for this ``candy'' diagram         
\begin{equation}\label{C1}
\begin{split}
\mathcal{FC}_{1}
=& \frac12 \left(\frac{i\lambda}{3\hbar H_0} \right)^2 \int\frac{d\omega_1 d\omega_2 d \omega_3 d \omega_4}{(2\pi)^4}\frac{e^{-i\omega_1t_1-i\omega_2t_2+i\omega_3t_3+i\omega_4t_4}\,\,\left(\frac{H_0^3\hbar}{4\pi^2}\right)^2\,i^2}{\left(\slashed{m}^2+\omega_1^2\right)\left(\slashed{m}^2+\omega_2^2\right)(\slashed{m}+i\omega_3)(\slashed{m}+i\omega_4)}\\
&\hspace{6cm}\times\mathcal{I}_{\mathcal{FC}_{1}}(\omega_1,\omega_3)\,\slashed{\delta}(\omega_3+\omega_4-\omega_1-\omega_2) \\
&= \frac12 \left(\frac{i\lambda}{3\hbar H_0} \right)^2 \int\frac{d\omega_1 d\omega_2 d \omega_3 d \omega_4}{(2\pi)^4}\frac{e^{-i\omega_1t_1-i\omega_2t_2+i\omega_3t_3+i\omega_4t_4}\,\,\left(\frac{H_0^3\hbar}{4\pi^2}\right)^2\,i^2 (\slashed{m}-i\omega_3)(\slashed{m}-i\omega_4)}{\left(\slashed{m}^2+\omega_1^2\right)\left(\slashed{m}^2+\omega_2^2\right)(\slashed{m}^2+\omega_3^2)(\slashed{m}^2+\omega_4^2)}\\
&\hspace{6cm}\times\mathcal{I}_{\mathcal{HC}_{1}}(\omega_1,\omega_3)\,\slashed{\delta}(\omega_3+\omega_4-\omega_1-\omega_2)
\end{split}
\end{equation}
where the loop integral is 
\begin{equation}
\begin{split}
\mathcal{I}_{\mathcal{FC}_{1}}(\omega_1,\omega_3) &=\left(\frac{H_0^3\hbar}{4\pi^2}\right)^2\int\frac{d\sigma}{2\pi}\frac{1}{\left(\slashed{m}^2+\sigma^2\right)\left(\slashed{m}^2+\left(\omega_1-\omega_3+\sigma\right)^2\right)}\\
&=\left(\frac{H_0^3\hbar}{4\pi^2}\right)^2\frac{1}{\slashed{m}}\frac{1}{4\slashed{m}^2+\left(\omega_1-\omega_3\right)^2} 
\end{split}
\end{equation}
The other three diagrams are obtained by shifting the placement of the $G$ lines. It is easy to see that the top right diagram ($\mathcal{HC}_2$) is obtained by replacing $\left( \slashed{m}-i\omega_3 \right) \rightarrow \left( \slashed{m}+i\omega_1\right)$ in the numerator of (\ref{C1}), the bottom left ($\mathcal{FC}_3$) by $\left( \slashed{m}-i\omega_3 \right)\left( \slashed{m}-i\omega_4 \right) \rightarrow \left( \slashed{m}+i\omega_1\right)\left( \slashed{m}+i\omega_2\right)$ and the bottom right ($\mathcal{FC}_4$) by $\left( \slashed{m}-i\omega_4 \right) \rightarrow \left( \slashed{m}+i\omega_2\right)$. Remarkably, adding up all the diagrams
results in cancellations leading to      

\begin{equation}
\label{horsum}
\sum\limits_i \mathcal{FC}_i (t_1,t_2,t_3,t_4)=\frac{1}{2\slashed{m}} \left(\frac{\lambda}{3\hbar \, H_0 }\right)^2 
\int \frac{\prod_i d\omega_i}{(2\pi)^4} \prod_i F(\omega_i)\, e^{i \Sigma \left(\omega_i t_i\right)} \slashed{\delta}(\Sigma \omega_i)	
\end{equation}
%
where we redefined the signs of all the frequencies, which is possible since $F(\omega)$ is even. 

The other three possibilities, connecting $t_1$ to $t_2$ or $t_1$ to $t_4$ without going through the loop, give an identical result and therefore all F-Candy diagrams contribute   
\begin{equation}
\mathcal{FC}(t_1,t_2,t_3,t_4)=\frac{3}{2\slashed{m}} \left(\frac{\lambda}{3\hbar \, H_0 }\right)^2 
\int \frac{\prod_i d\omega_i}{(2\pi)^4} \prod_i F(\omega_i)\, e^{i \Sigma \left(\omega_i t_i\right)} \slashed{\delta}(\Sigma \omega_i)	
\end{equation}

\subsubsection{$G$ Candies}
The other set of 1-loop diagrams contributing are $G$-candies in which one of the internal loop propagators is of $G$-type. Choosing the external times to be connected as seen below, 

\begin{table}[h]
	\begin{center}
		\begin{tabular}{cc} 
			
			\begin{tikzpicture}
			\filldraw[ultra thick] (0,1)circle (2pt) node[align=center, below]{$$};
			\filldraw[ultra thick] (0,0)circle (2pt) node[align=center, above]{$$}; 
			\filldraw[ultra thick] (-1,2)circle (2pt) node[align=center, below]{$t_1$};
			\filldraw[ultra thick] (1,2)circle (2pt) node[align=center, below]{$t_2$};
			\filldraw[ultra thick] (-1,-1)circle (2pt) node[align=center, below]{$t_3$};
			\filldraw[ultra thick] (1,-1)circle (2pt) node[align=center, below]{$t_4$};
			\filldraw[ultra thick](-1,2)--(-0.5,1.5);
			\filldraw[ultra thick](1,2)--(0,1);
			\filldraw[ultra thick](-1,-1)--(0,0);
			\filldraw[ultra thick](1,-1)--(0,0);
			\draw [ ultra thick] (0,0) arc (-90:90:0.5cm);
			\draw[ultra thick,decoration={snake,amplitude=0.06cm,segment length={\pathlen /5}},
			decorate] (0,0) arc (270:180:0.5cm);
			\draw [ ultra thick] (-0.5,0.5) arc (180:90:0.5cm);
			\draw[ultra thick, snake ] (0,1) -- (-0.5,1.5);
			\end{tikzpicture}
			&
			\begin{tikzpicture}
			\filldraw[ultra thick] (0,1)circle (2pt) node[align=center, below]{$$};
			\filldraw[ultra thick] (0,0)circle (2pt) node[align=center, above]{$$}; 
			\filldraw[ultra thick] (-1,2)circle (2pt) node[align=center, below]{$t_1$};
			\filldraw[ultra thick] (1,2)circle (2pt) node[align=center, below]{$t_2$};
			\filldraw[ultra thick] (-1,-1)circle (2pt) node[align=center, below]{$t_3$};
			\filldraw[ultra thick] (1,-1)circle (2pt) node[align=center, below]{$t_4$};
			\filldraw[ultra thick](-1,2)--(0,1);
			\filldraw[ultra thick](1,2)--(0.5,1.5);
			\filldraw[ultra thick](-1,-1)--(0,0);
			\filldraw[ultra thick](1,-1)--(0,0);
			\draw [ ultra thick] (0,0) arc (-90:90:0.5);
			\draw[ultra thick,decoration={snake,amplitude=0.06cm,segment length={\pathlen /5}},
			decorate] (0,0) arc (270:180:0.5cm);
			\draw [ ultra thick] (-0.5,0.5) arc (180:90:0.5cm);
			\draw[ultra thick, snake ] (0,1) -- (0.5,1.5);
			\end{tikzpicture}\\
			\begin{tikzpicture}
			\filldraw[ultra thick] (0,1)circle (2pt) node[align=center, below]{$$};
			\filldraw[ultra thick] (0,0)circle (2pt) node[align=center, above]{$$}; 
			\filldraw[ultra thick] (-1,2)circle (2pt) node[align=center, below]{$t_1$};
			\filldraw[ultra thick] (1,2)circle (2pt) node[align=center, below]{$t_2$};
			\filldraw[ultra thick] (-1,-1)circle (2pt) node[align=center, below]{$t_3$};
			\filldraw[ultra thick] (1,-1)circle (2pt) node[align=center, below]{$t_4$};
			\filldraw[ultra thick](-1,2)--(0,1);
			\filldraw[ultra thick](1,2)--(0,1);
			\filldraw[ultra thick](-1,-1)--(-0.5,-0.5);
			\filldraw[ultra thick](1,-1)--(0,0);
			\draw [ ultra thick] (0,0) arc (-90:90:0.5);
			\draw[ultra thick,decoration={snake,amplitude=0.06cm,segment length={\pathlen /5}},
			decorate] (-.5,.5) arc (180:90:.5);
			\draw [ ultra thick] (-.5,.5) arc (180:270:.5cm);
			\draw[ultra thick, snake ] (0,0) -- (-.5,-.5);
			\end{tikzpicture}
			&
			\begin{tikzpicture}
			\filldraw[ultra thick] (0,1)circle (2pt) node[align=center, below]{$$};
			\filldraw[ultra thick] (0,0)circle (2pt) node[align=center, above]{$$}; 
			\filldraw[ultra thick] (-1,2)circle (2pt) node[align=center, below]{$t_1$};
			\filldraw[ultra thick] (1,2)circle (2pt) node[align=center, below]{$t_2$};
			\filldraw[ultra thick] (-1,-1)circle (2pt) node[align=center, below]{$t_3$};
			\filldraw[ultra thick] (1,-1)circle (2pt) node[align=center, below]{$t_4$};
			\filldraw[ultra thick](-1,2)--(0,1);
			\filldraw[ultra thick](1,2)--(0,1);
			\filldraw[ultra thick](-1,-1)--(0,0);
			\filldraw[ultra thick](1,-1)--(0.5,-0.5);
			\draw [ ultra thick] (0,0) arc (-90:90:0.5);
			\draw[ultra thick,decoration={snake,amplitude=0.06cm,segment length={\pathlen /5}},
			decorate] (-.5,.5) arc (180:90:0.5cm);
			\draw [ ultra thick] (-.5,.5) arc (180:270:.5);
			\draw[ultra thick, snake ] (0,0) -- (.5,-.5);
			\end{tikzpicture}			
		\end{tabular}
	\end{center}
\end{table}

\noindent let us compute the top left diagram of this group. Assigning incoming $\omega_1$ and $\omega_3$ to $t_1$ and $t_3$ respectively, outgoing $\omega_2$ and $\omega_4$ to $t_2$ and $t_4$, frequency $\sigma$ running counter-clockwise in the loop, and noting that the symmetry factor is now unity (no possible exchange of $F$-lines in the loop), we have    
\begin{equation}
\begin{split}
\mathcal{GC}_1=& \, \frac{1}{2 \slashed{m}} \, \left[\frac{\lambda}{3\hbar \, H_0 }\right]^2 \int \frac{d\omega_1 \, d\omega_2 \, d\omega_3 \, d\omega_4 }{(2\pi)^{4}} \,  e^{-i\omega_1t_1 +i\omega_2t_2 -i\omega_3t_3 +i\omega_4t_4 }  \frac{\slashed{m}+i\omega_1}{(\slashed{2m})-i(\omega_1-\omega_2)} \\
&\cdot  F(\omega_1) \, F(\omega_2)\, F(\omega_3) \,F(\omega_4) \slashed{\delta}(-
\omega_1+\omega_2-\omega_3+\omega_4)
\end{split}
\end{equation}
where we directly included the loop integral  
\begin{equation}
\mathcal{I}_{\mathcal{GC}_1} = \int\frac{d\sigma}{2\pi}\frac{i}{\left(\slashed{m}-i\left(\sigma+\omega_1-\omega_2  \right)\right)}\frac{\frac{H_0^3\hbar}{4\pi^2}}{\slashed{m}^2+\sigma^2}
\end{equation} 
The other 3 diagrams are obtained by performing the appropriate permutations, as above, leading to a total of 
\begin{equation}
\begin{split}
\label{grsum}
\sum\limits_i\mathcal{G}=
&\, \frac{1}{2\slashed{m}} \, \left[\frac{\lambda}{3\hbar \, H_0 }\right]^2 
 \int \frac{d\omega_1 \, d\omega_2 \, d\omega_3 \, d\omega_4 }{(2\pi)^{4}} \,  e^{i( -\omega_1\, t_1 +\omega_2\, t_2 -\omega_3\, t_3 +\omega_4\, t_4 )} \prod_i F(\omega_i) \slashed{\delta}(-\omega_1+\omega_2-\omega_3+\omega_4)\\
&\times \left[ \frac{\slashed{m}+i\omega_1}{(\slashed{2m})+i(\omega_1-\omega_2)} +
              \frac{\slashed{m}-i\omega_2}{(\slashed{2m})+i(\omega_1-\omega_2)} + \frac{\slashed{m}+i\omega_3}{(\slashed{2m})-i(\omega_1-\omega_2)} + \frac{\slashed{m}-i\omega_4}{(\slashed{2m})-i(\omega_1-\omega_2)} \right] \\
=& \, \frac{1}{\slashed{m}} \, \left[\frac{\lambda}{3\hbar \, H_0 }\right]^2   \int \frac{\prod_i d\omega_i}{(2\pi)^4} \,  e^{i(\Sigma \omega_i\, t_i)} \prod_i F(\omega_i) \slashed{\delta}(\Sigma \omega_i) 
\end{split}
\end{equation}
where in the last equation the relevant $\omega_i\rightarrow-\omega_i, i=1,2$ transformations have been performed. As in the case of ${F}$- Candies, the other three possibilities, connecting $t_1$ to $t_2$ or $t_1$ to $t_4$ without going through the loop, give an identical result and therefore all $G$-Candy diagrams contribute   
 \begin{equation}
 \mathcal{GC}(t_1,t_2,t_3,t_4)=\frac{3}{\slashed{m}} \left(\frac{\lambda}{3\hbar \, H_0 }\right)^2 
 \int \frac{\prod_i d\omega_i}{(2\pi)^4} \prod_i F(\omega_i)\, e^{i \Sigma \left(\omega_i t_i\right)} \slashed{\delta}(\Sigma \omega_i)	
 \end{equation} 
Finally, adding the ${F}$- Candies and the ${G}$- Candies results in: 

\begin{equation}\begin{split}
\mathcal{Z}(t_1,t_2,t_3,t_4)&=\frac92 \left(\frac{\lambda}{3\hbar H_0}\right)^2 \,  \int \frac{\prod_i d\omega_i}{ (2\pi)^4} \,\cdot \slashed{\delta}(\Sigma \omega_i) \, e^{i(\Sigma \omega_i\, t_i)} \, \prod_i F(\omega_i)\\
&=18\frac{\hbar^2}{\slashed{m}}\frac{H_0^4}{\left(4\pi^2\right)^2}\left( \frac{\lambda H_0^2}{6\slashed{m}^24\pi^2}\right)^2\int \frac{\prod_i d\sigma_i}{ (2\pi)^4}  \prod_i \frac{1}{1+\sigma_i^2}\,\cdot \slashed{\delta}(\Sigma \sigma_i) \, e^{i(\slashed{m}\Sigma \sigma_i\, t_i)}
\end{split}
\end{equation}
Again, the expansion parameter is $\frac{\lambda H_0^2}{6\slashed{m}^24\pi^2}$, consistent with the 2-point function,

\section{Stochastic diagrams from the Langevin equation in Pure deSitter}\label{sec:direct-Langevin}
In this section we obtain the 2-point and 4-point functions to order $\lambda^2$ directly from the Langevin equation for the potential (\ref{potential}). We show how this perturbative solution can be represented graphically and how we can eventually obtain diagrams that end up being identical to the Feynman diagrams of section \ref{sect:path integral}. However, obtaining them needs a non-insignificant amount of labour compared to the direct application of the Feynman rules stated in the previous section. Therefore, this section not only provides a check of the previous computations but also demonstrates the efficiency of using the Feynman rules stated in section \ref{sect:path integral} compared to working with the direct solution of the Langevin equation.           

Let us again write down the Langevin equation 
\begin{equation}
\label{equation1}
\dot{\phi}+\slashed{m} \phi+\frac{\lambda}{6  \cdot 3\hbar \, H_0}\phi^3=\hbar^{1/2}\mathcal{A}\,\xi(t)
\end{equation}
where, for simplicity we ignore in this section any dependence of $\mathcal{A}$ on $\phi$. Expanding the solution  $\phi(t)=\phi_{(0)}+\phi_{(1)}+\phi_{(2)}+\ldots$ to different orders in $\lambda$ and accordingly splitting the Langevin equation \eqref{equation1} into equations of different orders results in:

\begin{eqnarray}
\label{orderlambda0}
\dot{\phi}_{(0)} + \slashed{m} \phi_{(0)}&=&\hbar^{1/2}\mathcal{A}\,\xi(t)\\ 
\label{orderlambda1}
\dot{\phi}_{(1)} + \slashed{m} \phi_{(1)} + \frac{\lambda}{6  \cdot 3\hbar \, H_0} \phi^3_{(0)}&=& 0\\
\label{orderlambda2}
\dot{\phi}_{(2)} + \slashed{m} \phi_{(2)} + \frac{\lambda}{6  \cdot 3\hbar \, H_0} \left(3 \phi_{(0)}^2  \phi_{(1)}\right)&=& 0 \\
&\vdots\nonumber
\end{eqnarray}
By defining the Fourier transform as in (\ref{Fourier}), we directly obtain in Fourier space and to order $O(\lambda^0)$  
\begin{equation}
\label{Phi0}
\phi_{(0)}(\omega) =\frac{1}{{\slashed{m}}+i\omega} \hbar^{1/2}\mathcal{A}\,\xi(\omega) \equiv G^R(\omega)\hbar^{1/2}\mathcal{A}\,\xi(\omega). 
\end{equation} 
Fourier transforming the cubic term in \eqref{orderlambda1}  
\begin{equation}
\phi^3_{(0)}(t)=\int\frac{d\omega \, d\omega' \, d\omega''}{{(2\pi)}^3} \  \phi_{(0)}(\omega) \, \phi_{(0)}(\omega') \,  \phi_{(0)}(\omega'') \, e^{i(\omega+\omega'+\omega'')t}
\end{equation}
the first order equation reads in Fourier space 
\begin{equation}
\begin{split}
\label{x1calculation}
\left(\slashed{m} +i \omega \right)& \phi_{(1)}(\omega) \\ 
=&- \frac{\lambda }{6  \cdot 3\hbar \, H_0 } \int \frac{d\omega_1 \, d\omega_2  \,d\omega_3}{(2\pi)^3}
\slashed{\delta}(\omega_1+\omega_2+\omega_3 -\omega) \phi_0({\omega_1}) \phi_0({\omega_2}) \phi_0({\omega_3})\\
=&- \frac{\lambda \, \left( \hbar^{\frac12} \mathcal{A} \right)^3 }{6  \cdot 3\hbar \, H_0 } \int \frac{d\omega_1 \, d\omega_2  \,d\omega_3}{(2\pi)^3}
\slashed{\delta}(\omega_1+\omega_2+\omega_3 -\omega) \, G^R(\omega_1) \xi(\omega_1) \,  G^R(\omega_2) \xi(\omega_2) \, G^R(\omega_3)\xi(\omega_3).
\end{split}
\end{equation}
and therefore the $O(\lambda^1)$ solution is straightforwardly obtained  
\begin{equation}
\label{x1}
\phi_{(1)}(\omega) =- \frac{\lambda \ \left(\hbar^{1/2}\mathcal{A} \right)^3}{6  \cdot 3\hbar \, H_0 } G^R( \omega ) \int \frac{d\omega_1 \, d\omega_2  \,d\omega_3}{(2\pi)^3}
\slashed{\delta}(\omega_1+\omega_2+\omega_3 -\omega) \, G^R(\omega_1)\xi(\omega_1) \,  G^R(\omega_2)\xi(\omega_2) \, G^R(\omega_3)\xi(\omega_3).
\end{equation}
In a similar vain, the $O(\lambda^2)$ solution reads 

\begin{equation}
\begin{split}
\label{x2}
\phi_{(2)}(\omega) &= \frac{\lambda^2 \, \left(\hbar^{1/2}\mathcal{A} \right)^5}{12 \hbar^2 \cdot 9H_0^2} G^R(\omega)\\
&  \times \int \frac{d\omega_1 \, d\omega_2 \, d\omega_3 \, d\omega_1' \, d\omega_2' \, d\omega_3'}{(2\pi)^6}  \slashed{\delta}(\omega_1+\omega_2+\omega_3-\omega_3') \, \slashed{\delta}(\omega_1'+\omega_2'+\omega_3'-\omega) \\	
& \times
G^R(\omega_1')\xi(\omega_1')\
G^R(\omega_2')\xi(\omega_2')\
G^R(\omega_3')              \
G^R(\omega_1) \xi(\omega_1)\
G^R(\omega_2) \xi(\omega_2)\
G^R(\omega_3) \xi(\omega_3)\
\end{split}.
\end{equation}
These results can be represented in a graphical way as the tree graphs seen below 
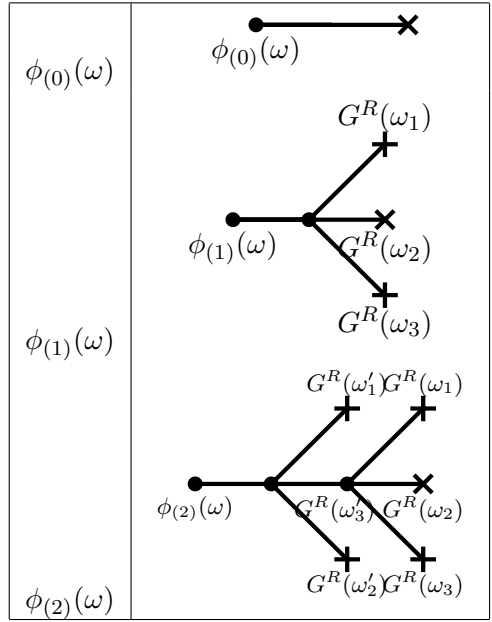
\begin{center}
	\begin{tabular}{ |c|c| } 
		\hline
		$\phi_{(0)}(\omega)$ 
		& 	
		\begin{tikzpicture}[]
		\vspace{3cm}\draw[ultra thick] (1,0)  node[cross=5pt,rotate=90] {} ;
		\filldraw[ultra thick] (1,0) node[]{} -- (-1,0)circle (2pt) node[align=center, below] {$\phi_{(0)}(\omega)$} -- (0,0)circle (0pt) node[align=right, above]{}; 
		\end{tikzpicture}   \\ 
		$\phi_{(1)}(\omega)$ &
		\begin{tikzpicture}
		\draw[ultra thick] (1,1)  node[cross=5pt,rotate=45] {} ;
		\draw[ultra thick] (1,0)  node[cross=5pt,rotate=90] {} ;
		\draw[ultra thick] (1,-1)  node[cross=5pt,rotate=-45] {} ;
		\filldraw[ultra thick] (1,0)circle (0pt) node[align=center, below] {$G^R(\omega_2)$} -- (-1,0)circle (2pt) node[align=center, below] {$\phi_{(1)}(\omega)$} -- (0,0)circle (2pt) node[align=right, above] {} -- (1,1)circle (0pt) node[align=center, above] {$G^R(\omega_1)$} ; 
		\filldraw[ultra thick] (0,0) -- (1,-1)circle (0pt) node[align=center, below] {$G^R(\omega_3)$};
		\end{tikzpicture} 	\\ 
		$\phi_{(2)}(\omega)$ &
		\begin{tikzpicture}
		\draw[ultra thick] (1,1)  node[cross=5pt,rotate=45] {} ;
		\draw[ultra thick] (1,-1)  node[cross=5pt,rotate=45] {} ;
		\draw[ultra thick] (2,1)  node[cross=5pt,rotate=45] {} ;
		\draw[ultra thick] (2,0)  node[cross=5pt,rotate=0] {} ;
		\draw[ultra thick] (2,-1)  node[cross=5pt,rotate=-45] {} ;
		\filldraw[ultra thick] (-1,0)circle (2pt) node[align=center, below] {\footnotesize $\phi_{(2)}(\omega)$} -- (1,0)circle (2pt) node[align=left, below] {\footnotesize $G^R(\omega'_3) \,\,\,\,$ };
		\filldraw[ultra thick] (0,0)circle (2pt) node[align=right, above]{} -- (1,1)circle (0pt) node[align=center, above] {\footnotesize $G^R(\omega'_1)$} ; 
		\filldraw[ultra thick] (0,0) -- (1,-1)circle (0pt) node[align=center, below] {\footnotesize$G^R(\omega'_2)$};
		\filldraw[ultra thick] (1,0) -- (2,1)circle (0pt) node[align=center, above] {\footnotesize$G^R(\omega_1)$};
		\filldraw[ultra thick] (1,0) -- (2,0)circle (0pt) node[align=center, below] {\footnotesize$G^R(\omega_2)$};
		\filldraw[ultra thick] (1,0) -- (2,-1)circle (0pt) node[align=center, below] {\footnotesize$G^R(\omega_3)$}; 
		\end{tikzpicture} \\
		\hline
	\end{tabular}
\end{center}
Crosses represent $\xi$ sources while lines stand for retarded propagators $G^R$ which evolve the sources to build up the field or, equivalently, its Fourier transform at frequency $\omega$. Note that  different crosses can be thought of as injecting different frequencies in the tree and each vertex conserves the total frequency flowing in and out of it. Higher orders can be obtained similarly, as increasingly complex trees.      

\subsection{Two-point function}
Let us look again at the two-point function
\begin{equation}
\left\langle\phi(\omega)\phi(\omega')\right\rangle
\end{equation}
up to 2nd order in $\lambda$. Expanding as above we have  
\begin{equation}
\begin{split}
\label{propsplit}
\left\langle\phi(\omega)\phi(\omega')\right\rangle=&
\left\langle\left(\phi_{(0)}(\omega)+\phi_{(1)}(\omega)+\phi_{(2)}(\omega) + \ldots\right)\left(\phi_{(0)}(\omega')+\phi_{(1)}(\omega')+\phi_{(2)}(\omega')+\ldots\right)\right\rangle \\
=&\left\langle\phi_{(0)}(\omega)\phi_{(0)}(\omega')\right\rangle\\
+&\left\langle\phi_{(0)}(\omega)\phi_{(1)}(\omega')\right\rangle +\left\langle\phi_{(1)}(\omega)\phi_{(0)}(\omega')\right\rangle\\
+&\left\langle\phi_{(0)}(\omega)\phi_{(2)}(\omega')\right\rangle + \left\langle\phi_{(2)}(\omega)\phi_{(0)}(\omega')\right\rangle + \left\langle\phi_{(1)}(\omega)\phi_{(1)}(\omega')\right\rangle + \ldots
\end{split}
\end{equation}
Substituting from \eqref{Phi0}, the correlator to 0th order in $\lambda$ is given as 
\begin{equation}
\left\langle\phi_{(0)}(\omega)\phi_{(0)}(\omega')\right\rangle 
=  \frac{\hbar \, H_0^3}{4\pi^2}\frac{1}{{\slashed{m}^2}+\omega^2} \slashed{\delta}(\omega+\omega'), 
\end{equation} 
where the Fourier transformation of $\xi$ and \eqref{xisquared}
have been used to obtain 
\begin{equation}
\left\langle{\xi}(\omega){\xi}(\omega')\right\rangle = \slashed{\delta}(\omega+\omega').
\end{equation}
The operation of taking the expectation value on the product of noise terms can be graphically represented as 

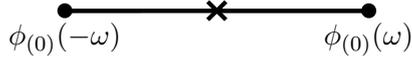
\begin{tikzpicture}[]
\vspace{3cm}\draw[ultra thick] (1,0)  node[cross=5pt,rotate=90] {} ;
\filldraw[ultra thick] (1,0) node[]{} -- (-1,0)circle (2pt) node[align=center, below] {$\phi_{(0)}(-\omega)$} -- (0,0)circle (0pt) node[align=right, above]{};
\filldraw[ultra thick] (1,0) node[]{} -- (3,0)circle (2pt) node[align=center, below] {$\phi_{(0)}(\omega)$} -- (0,0)circle (0pt) node[align=right, above]{};
\end{tikzpicture}  

\noindent i.e. to obtain correlators two crosses can be joined together producing an $F$-line with frequency $\omega$ flowing across it.

\subsubsection{$O(\lambda)$}
For the {$O(\lambda^1)$} contribution to the correlator we have  

\begin{equation}
\begin{split}
\label{x0x1}
\left\langle{\phi}_{(0)}(t){\phi}_{(1)}(t')\right\rangle\, 
&=\int \frac{d\omega \, d\omega'}{\left(2 \pi \right)^2 } e^{i(\omega t +\omega' t')} \left\langle\phi_{(0)}(\omega)\phi_{(1)}(\omega')\right\rangle  \\
&=\left(\frac{\hbar H_0^3}{4\pi^2}\right)^2   \int \frac{d\omega_1\,d\omega_2 \, d\omega_3\, d\omega\,d\omega'}{(2\pi)^5}  e^{i(\omega t +\omega' t')} \left(-\frac{\lambda}{6\hbar \cdot 3H_0} G^R(\omega) \,  G^R(\omega')\right)\\
&\hspace{1.5cm}\times \Big\{\left\langle\xi(\omega)\xi(\omega_1)\right\rangle \left\langle\xi(\omega_2)\xi(\omega_3)\right\rangle \\
&  \hspace{2cm}+ \left\langle\xi(\omega)\xi(\omega_2)\right\rangle \left\langle\xi(\omega_1)\xi(\omega_3)\right\rangle \\
&  \hspace{2cm}+ \left\langle\xi(\omega)\xi(\omega_3)\right\rangle \left\langle\xi(\omega_1)\xi(\omega_2)\right\rangle \Big\}\\
&\hspace{1.5cm} \times G^R(\omega_1) G^R(\omega_2)G^R(\omega_3) \,\,\slashed{\delta}(\omega_1+\omega_2 +\omega_3-\omega') 
\end{split}
\end{equation}
where Wick's theorem was used to expand $\left\langle\xi(\omega)\xi(\omega_1)\xi(\omega_2)\xi(\omega_3)\right\rangle$. Thus, 
\begin{equation}
\left\langle{\phi}_{(0)}(t){\phi}_{(1)}(t')\right\rangle = -\frac{\lambda H_0^5 \, \hbar }{\left(8 \pi^2\right)^2 \,3  \slashed{m} } \int \frac{d\omega}{2\pi} \frac{1}{\slashed{m}^2+\omega^2} \frac{1}{\slashed{m}-i\,\omega} e^{i\omega(t-t')}
\end{equation}
and the Fourier space correction to the two-point function is 

\begin{equation}
\label{x0x1fourier}
\Delta(\omega)=-\frac{\lambda H_0^5 \, \hbar}{\left(8 \pi^2\right)^2 \,3 \slashed{m} }   \frac{1}{\slashed{m}^2+\omega^2} \frac{1}{\slashed{m}-i\,\omega}.
\end{equation}

In order to obtain the term, $\left\langle{\phi}_{(0)}(t'),{\phi}_{(1)}(t)\right\rangle$, one may observe that changing $t\leftrightarrow t'$ and $\omega\leftrightarrow \omega'$ in equation \eqref{x0x1}, results in the expression in question:
\begin{equation}
\label{x1x0}
\left\langle{\phi}_{(0)}(t'),{\phi}_{(1)}(t)\right\rangle=-\frac{\lambda H_0^5\, \hbar}{\left(8 \pi^2\right)^2 \,3  \slashed{m} } \int \frac{d\omega}{2\pi} \frac{1}{\slashed{m}^2+\omega^2} \frac{1}{\slashed{m}+i\,\omega} e^{i\omega(t-t')}
\end{equation}
Hence, in Fourier space: 
\begin{equation}
\label{x1x0fourier}
\Delta'(\omega)=-\frac{\lambda H_0^5\, \hbar}{\left(8 \pi^2\right)^2 \,3  \slashed{m} }  \frac{1}{\slashed{m}^2+\omega^2} \frac{1}{\slashed{m}+i\,\omega}.
\end{equation}
Finally, adding \eqref{x0x1fourier} and \eqref{x1x0fourier} results in the $O(\lambda)$ contribution to the two-point function
\begin{equation}
\label{fourierprop1}
F_1(\omega) =-\frac{2 \lambda H_0^5\, \hbar}{3\left(8 \pi^2\right)^2  } \frac{1}{\left[ \slashed{m}^2 +\omega^2 \right]^2}
\end{equation}
which, as expected, is exactly equivalent to \eqref{propagatorl1}.
This result can be obtained graphically by joining all the crosses in the trees representing $\phi_{(0)}$ and $\phi_{(1)}$.      

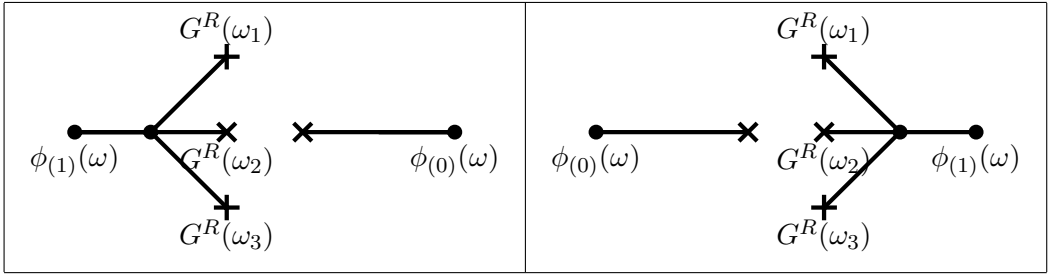
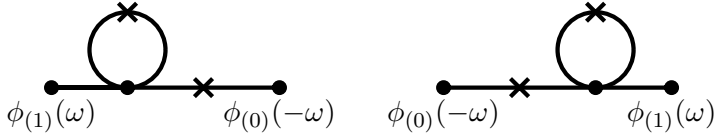
\begin{center}
	\begin{tabular}{ |c|c| } 
		\hline
		
	\begin{tikzpicture}
		\draw[ultra thick] (2,0)  node[cross=5pt,rotate=90] {} ;
		\filldraw[ultra thick] (3,0) node[]{} -- (4,0)circle (2pt) node[align=center, below] {$\phi_{(0)}(\omega)$} -- (2,0)circle (0pt) node[align=right, above]{}; 
		
		\draw[ultra thick] (1,1)  node[cross=5pt,rotate=45] {} ;
		\draw[ultra thick] (1,0)  node[cross=5pt,rotate=90] {} ;
		\draw[ultra thick] (1,-1)  node[cross=5pt,rotate=-45] {} ;
		\filldraw[ultra thick] (1,0)circle (0pt) node[align=center, below] {$G^R(\omega_2)$} -- (-1,0)circle (2pt) node[align=center, below] {$\phi_{(1)}(\omega)$} -- (0,0)circle (2pt) node[align=right, above] {} -- (1,1)circle (0pt) node[align=center, above] {$G^R(\omega_1)$} ; 
		\filldraw[ultra thick] (0,0) -- (1,-1)circle (0pt) node[align=center, below] {$G^R(\omega_3)$};
		\end{tikzpicture}
		& 	
		\begin{tikzpicture}
		\draw[ultra thick] (1,0)  node[cross=5pt,rotate=90] {} ;
		\filldraw[ultra thick] (-1,0)circle (2pt) node[align=center, below] {$\phi_{(0)}(\omega)$} -- (1,0)circle (0pt)  ; 
		
		\draw[ultra thick] (2,1)  node[cross=5pt,rotate=45] {} ;
		\draw[ultra thick] (2,0)  node[cross=5pt,rotate=90] {} ;
		\draw[ultra thick] (2,-1)  node[cross=5pt,rotate=-45] {} ;
		\filldraw[ultra thick] (4,0)circle (2pt) node[align=center, below] {$\phi_{(1)}(\omega)$} -- (3,0)circle (2pt) node[align=center, below] {};
		\filldraw[ultra thick](3,0) node[align=right] {} -- (2,1)circle (0pt) node[align=center, above] {$G^R(\omega_1)$} ; 
		\filldraw[ultra thick] (3,0) -- (2,-1)circle (0pt) node[align=center, below] {$G^R(\omega_3)$};
		\filldraw[ultra thick] (3,0) -- (2,0)circle (0pt) node[align=center, below] {$G^R(\omega_2)$};
		\end{tikzpicture}  \\ 
		\hline
	\end{tabular}
\end{center}

as seen below

\vspace{0.3cm}

\begin{tabular}{ c c } 
		\begin{tikzpicture}
		\draw[ultra thick] (1,0)  node[cross=5pt,rotate=90] {} ;
		\filldraw[ultra thick] (2,0) circle (2pt) node[align=center, below] {$\phi_{(0)}(-\omega)$}  -- (1,0);
		\filldraw[ultra thick] (1,0)circle (0pt) node[align=center, below] {} -- (-1,0)circle (2pt) node[align=center, below] {$\phi_{(1)}(\omega)$} -- (0,0)circle (2pt) node[align=right, above] {} ;
		\draw [ ultra thick] (0.5,0.5) arc (0:360:0.5cm);
		\draw[ultra thick] (0,1)  node[cross=5pt,rotate=90] {} ; 
		\end{tikzpicture}
		& 	
		\begin{tikzpicture}
		\filldraw[ultra thick] (1,0)circle (2pt) node[align=center, below] {$\phi_{(0)}(-\omega)$} -- (2,0)circle (0pt)  ; 
		\draw[ultra thick] (2,0)  node[cross=5pt,rotate=90] {} ;
		\filldraw[ultra thick] (4,0)circle (2pt) node[align=center, below] {$\phi_{(1)}(\omega)$} -- (3,0)circle (2pt) node[align=center, below] {};
		\filldraw[ultra thick] (3,0) -- (2,0)circle (0pt) node[align=center, below] {};
		\draw [ ultra thick] (3.5,0.5) arc (0:360:0.5cm);
		\draw[ultra thick] (3 , 1)  node[cross=5pt,rotate=90] {} ;
		\end{tikzpicture}  \\ 
	\end{tabular}

\noindent One can easily see that the resulting diagrams are equivalent to those obtained in section \ref{sect:path integral} directly using the Feynman rules by noting that a crossed line here equates to a straight $F$-line and a straight line here equates to a straight-jagged line in the path integral formalism. 

\vspace{0.3cm}
	
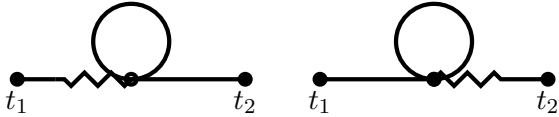
\begin{tabular}{ c c } 
	
	\begin{tikzpicture}
	\filldraw [ultra thick](1.5,0)circle (2pt) node[align=center, below] {$t_2$} -- (0,0);
	\draw [  snake,ultra thick   ] (0,0)circle (2pt) node[align=center, below]{$$} -- (-1,0);
	\filldraw [ultra thick] (-1.5,0)circle (2pt) node[align=center, below] {$t_1$}--(-1,0);
	\draw [ultra thick] (0,0.5) circle [radius=0.5];
	\end{tikzpicture}
	&
	\begin{tikzpicture}
	\filldraw[ultra thick] (-1.5,0)circle (2pt) node[align=center, below] {$t_1$} -- (0,0)circle (2pt);
	\draw [  snake, ultra thick  ] (0,0)circle (2pt) node[align=center, below]{$$} -- (1,0);
	\filldraw[ultra thick] (1,0)--(1.5,0)circle (2pt) node[align=center, below] {$t_2$};
	\draw [ ultra thick] (0,0.5) circle [radius=0.5];
	\end{tikzpicture} \\
	
\end{tabular}

\noindent This correspondence is realised to all orders and for all diagrams.

\subsubsection{$O(\lambda^2)$}

The terms that contribute to $O(\lambda^2)$ can be seen in \eqref{propsplit} to be $\left\langle\phi_{(0)}(\omega)\phi_{(2)}(\omega')\right\rangle$,  $\left\langle\phi_{(2)}(\omega)\phi_{(0)}(\omega')\right\rangle$ and  $\left\langle\phi_{(1)}(\omega)\phi_{(1)}(\omega')\right\rangle$. Having calculated $\phi_{(0)}(\omega)$,$\phi_{(1)}(\omega)$ and $\phi_{(2)}(\omega)$ in \eqref{Phi0}, \eqref{x1} and \eqref{x2} respectively we determine the NNLO correction to the two-point function as follows:
\begin{equation}
\begin{split}
\label{x1x1}
\left\langle\phi_{(1)}(\omega)\phi_{(1)}(\omega')\right\rangle&=\frac{\lambda^2}{(6\hbar \cdot 3H_0^2)}  \left(\frac{\hbar H_0^3}{4\pi^2}\right)^3  G^R(\omega)G^R(\omega')\\
&\hspace{0.5cm}\cdot\int   \frac{d\omega_1d\omega_1d\omega_1d\omega_1' d\omega_2 'd\omega_3'}{(2\pi)^6} \slashed{\delta}(\omega_1+\omega_2+\omega_3-\omega)  \, \slashed{\delta}(\omega_1'+\omega_2'+\omega_3'-\omega')\\
&\hspace{0.5cm}\times G^R(\omega_1)G^R(\omega_2)G^R(\omega_3)G^R(\omega_1')G^R(\omega_2')G^R(\omega_3')\\
&\hspace{0.5cm} \times \left\langle\xi(\omega_1)\cdot\xi(\omega_2)\cdot\xi(\omega_3)\cdot\xi(\omega_1')\cdot\xi(\omega_2')\cdot\xi(\omega_3')\right\rangle
\end{split}
\end{equation} 

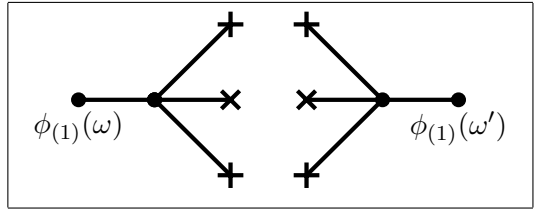
\begin{center}
	\begin{tabular}{ |c| } 
		\hline
		
		\begin{tikzpicture}
		\draw[ultra thick] (1,0)  node[cross=5pt,rotate=90] {} ;
		\draw[ultra thick] (1,1)  node[cross=5pt,rotate=45] {} ;
		\draw[ultra thick] (1,-1)  node[cross=5pt,rotate=45] {} ;
		\filldraw[ultra thick] (-1,0)circle (2pt) node[align=center, below] {$\phi_{(1)}(\omega)$} -- (0,0)circle (2pt) --(1,0)circle(0pt) ;
		\filldraw[ultra thick] (0,0)circle (2pt) --(1,1)circle(0pt) ;
		\filldraw[ultra thick] (0,0)circle (2pt) --(1,-1)circle(0pt) ;
		\draw[ultra thick] (2,1)  node[cross=5pt,rotate=45] {} ;
		\draw[ultra thick] (2,0)  node[cross=5pt,rotate=90] {} ;
		\draw[ultra thick] (2,-1)  node[cross=5pt,rotate=-45] {} ;
		\filldraw[ultra thick] (4,0)circle (2pt) node[align=center, below] {$\phi_{(1)}(\omega')$} -- (3,0)circle (2pt) node[align=center, below] {};
		\filldraw[ultra thick](3,0) node[align=right] {} -- (2,1)circle (0pt) node[align=center, above] {} ; 
		\filldraw[ultra thick] (3,0) -- (2,-1)circle (0pt) node[align=center, below] {};
		\filldraw[ultra thick] (3,0) -- (2,0)circle (0pt) node[align=center, below] {};
		\end{tikzpicture} 
		\\ 
		\hline
	\end{tabular}
\end{center}

Using Wick's theorem, it is evident that there are 15 terms of different pairs in  $\big\langle\tilde{\xi}(\omega_1)\cdot\tilde{\xi}(\omega_2)\cdot\xi(\omega_3)\cdot\xi(\omega_1')\cdot\xi(\omega_2')\cdot\xi(\omega_3')\big\rangle$. The diagrammatic presentation of those (or alternatively, the symmetries of the integrals and the delta functions) demonstrate that there are only two topologically inequivalent ways for these 15 terms to be organised:
6 "Symmetric Sunset" diagrams and 9 "Symmetric Double Seagull" diagrams.

\textbf{The Symmetric Sunset}

\begin{minipage}{3cm}
	\begin{tikzpicture}
	\filldraw[ultra thick](4,0)circle (2pt) node[align=center, below] {$t_2$} ;
	\filldraw[ultra thick](3,0)circle (2pt) node[align=center, below] {$\hspace{0.3cm}  $} ;
	\filldraw[ultra thick](2,0)circle (2pt) node[align=center, below] {$\hspace{-0.3cm} $} ;
	\filldraw[ultra thick](1,0)circle (2pt) node[align=center, below] {$t_1$} ;
	\filldraw [ultra thick](1,0) -- (2,0);
	\filldraw [ultra thick](2,0) -- (3,0);
	\filldraw [ultra thick](3,0) -- (4,0);
	\draw [ultra thick] (2.5,0) circle [radius=0.5];
	\draw[ultra thick] (2.5,0.5)  node[cross=5pt,rotate=0] {} ;
	\draw[ultra thick] (2.5,0)  node[cross=5pt,rotate=0] {} ;
	\draw[ultra thick] (2.5,-0.5)  node[cross=5pt,rotate=0] {} ;
	\end{tikzpicture}
\end{minipage}
\begin{minipage}{10cm}
	\begin{equation}
	\begin{split}
	\mathcal{SS}&=6 \left(\frac{  \lambda }{6\hbar \cdot 3H_0} \right)^2 \left( \frac{\hbar H_0^3}{4\pi^2} \right)^3 G^R(\omega) G^R(\omega') \int \frac{d\omega_1 d\omega_2 d\omega_3 d\omega'_1 d\omega'_2 d\omega'_3 }{(2\pi)^6}\\  &\hspace{0.5cm}\,\slashed{\delta}(\omega_1+\omega_2+\omega_3-\omega) \slashed{\delta}(\omega_1'+\omega_2'+\omega_3'-\omega')\\
	&\hspace{0.5cm}\cdot  G^R(\omega_1) \,G^R(\omega_2) \,G^R(\omega_3) \,G^R(\omega_1') \,G^R(\omega_2') \,G^R(\omega_3') \\
	&\hspace{0.5cm}\cdot \slashed{\delta}(\omega_1+\omega_1')\slashed{\delta}(\omega_2+\omega_2')\slashed{\delta}(\omega_3+\omega_3')\\
	&\hspace{0.5cm}=\frac{ \lambda^2 H_0^7 \, \hbar }{9\slashed{m}^2 \,\, \, \left(8\pi^2\right)^3 }  \frac{1 }{\left(\slashed{m}^2+\omega^2\right) \left[ \left(3\slashed{m}\right)^2 +\omega^2\right] },
	\end{split}
	\end{equation}
\end{minipage}

in exact agreement with \eqref{symsun}.

\textbf{The Symmetric Double Seagull}

\begin{minipage}{5cm}
	\begin{tikzpicture}
	\filldraw[ultra thick](4.5,0)circle (2pt) node[align=center, below] {$t_2$} ;
	\filldraw[ultra thick](3.5,0)circle (2pt) node[align=center, below] {$$} ;
	\filldraw[ultra thick](2,0)circle (2pt) node[align=center, below] {$$} ;
	\filldraw[ultra thick](1,0)circle (2pt) node[align=center, below] {$t_1$} ;
	\filldraw [ultra thick](1,0) -- (4.5,0);
	\draw [ultra thick] (2,0.5) circle [radius=0.5];
	\draw [ultra thick] (3.5,0.5) circle [radius=0.5];
	\draw[ultra thick] (2,1)  node[cross=5pt,rotate=0] {} ;
	\draw[ultra thick] (2.75,0)  node[cross=5pt,rotate=0] {} ;
	\draw[ultra thick] (3.5,1)  node[cross=5pt,rotate=0] {} ;
	\end{tikzpicture}
\end{minipage}
\begin{minipage}{10cm}
	\begin{equation}
	\begin{split}
	\mathcal{SS}&=9 \left(\frac{  \lambda }{6\hbar \cdot 3H_0} \right)^2 \left( \frac{\hbar H_0^3}{4\pi^2} \right)^3 G^R(\omega) G^R(\omega') \int \frac{d\omega_1 d\omega_2 d\omega_3 d\omega'_1 d\omega'_2 d\omega'_3 }{(2\pi)^6}\\  &\hspace{0.5cm}\,\slashed{\delta}(\omega_1+\omega_2+\omega_3-\omega) \slashed{\delta}(\omega_1'+\omega_2'+\omega_3'-\omega')\\
	&\hspace{0.5cm}\cdot  G^R(\omega_1) \,G^R(\omega_2) \,G^R(\omega_3) \,G^R(\omega_1') \,G^R(\omega_2') \,G^R(\omega_3') \\
	&\hspace{0.5cm}\cdot \slashed{\delta}(\omega_1+\omega_2)\slashed{\delta}(\omega_3+\omega_3')\slashed{\delta}(\omega_1'+\omega_2')\\
	&\hspace{0.5cm}=\frac{ \lambda^2 H_0^7 \, \hbar }{18\slashed{m}^2 \,\, \, \left(8\pi^2\right)^3 }  \frac{1 }{\left(\slashed{m}^2+\omega^2\right)}
	\end{split}
	\end{equation}
\end{minipage}

in exact agreement with \eqref{symsseagull}.

Furthermore, there are two more second-order in $\lambda$ contributions : 

\begin{equation}
\begin{split}
\label{phi2phi0}
\left\langle\phi_{(2)}(\omega')\phi_{(0)}(\omega)\right\rangle
=	&G^R(\omega')\left[\frac{\lambda^2  }{12\hbar^2 \cdot 3H_0^2}\right] \left[\frac{\hbar H_0^3}{4\pi^2}\right]^3 G^R(\omega) \\
&\int \frac{d\omega_1d\omega_2d\omega_3d\omega_1' d\omega_2 'd\omega_3'}{2\pi^6} \slashed{\delta}(\omega_1+\omega_2+\omega_3-\omega_3') \slashed{\delta}(\omega_1'+\omega_2'+\omega_3'-\omega)\\	&G^R(\omega_1)G^R(\omega_2)G^R(\omega_3)G^R(\omega_1')G^R(\omega_2')G^R(\omega_3')		
\end{split}
\end{equation}

and its $\omega \rightarrow \omega'$ symmetric.

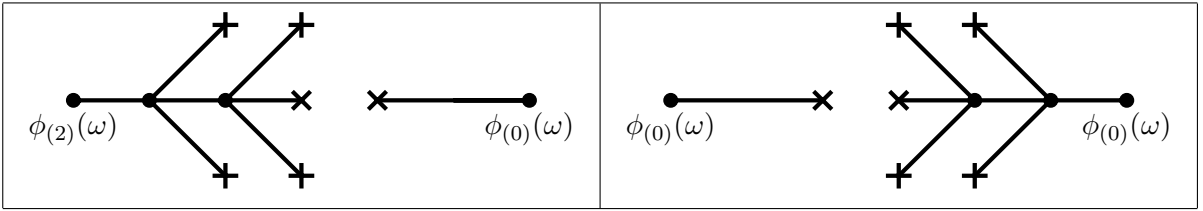
\begin{center}
	\begin{tabular}{ |c|c| } 
		\hline
		
		\begin{tikzpicture}
		\draw[ultra thick] (1,1)  node[cross=5pt,rotate=45] {} ;
		\draw[ultra thick] (1,-1)  node[cross=5pt,rotate=45] {} ;
		\draw[ultra thick] (2,1)  node[cross=5pt,rotate=45] {} ;
		\draw[ultra thick] (2,0)  node[cross=5pt,rotate=0] {} ;
		\draw[ultra thick] (2,-1)  node[cross=5pt,rotate=-45] {} ;
		\filldraw[ultra thick] (-1,0)circle (2pt) node[align=center, below] {$\phi_{(2)}(\omega)$} -- (1,0)circle (2pt) node[align=left, below] {};
		\filldraw[ultra thick] (0,0)circle (2pt) node[align=right,]{} -- (1,1)circle (0pt) node[align=center, above] {} ; 
		\filldraw[ultra thick] (0,0) -- (1,-1)circle (0pt) node[align=center, below] {};
		\filldraw[ultra thick] (1,0) -- (2,1)circle (0pt) node[align=center, above] {};
		\filldraw[ultra thick] (1,0) -- (2,0)circle (0pt) node[align=center, below] {};
		\filldraw[ultra thick] (1,0) -- (2,-1)circle (0pt) node[align=center, below] {}; 
		
		\draw[ultra thick] (3,0)  node[cross=5pt,rotate=90] {} ;
		\filldraw[ultra thick] (4,0) node[]{} -- (5,0)circle (2pt) node[align=center, below] {$\phi_{(0)}(\omega)$} -- (3,0)circle (0pt) node[align=right, above]{}; 
		
		\end{tikzpicture}
		& 	
		\begin{tikzpicture}
		\draw[ultra thick] (0,0)  node[cross=5pt,rotate=90] {} ;
		\filldraw[ultra thick] (-2,0) circle (2pt) node[align=center, below] {$\phi_{(0)}(\omega)$} -- (0,0)circle (0pt) node[align=right, above]{};

		\draw[ultra thick] (1,1)  node[cross=5pt,rotate=45] {} ;
		\draw[ultra thick] (1,-1)  node[cross=5pt,rotate=45] {} ;
		\draw[ultra thick] (2,1)  node[cross=5pt,rotate=45] {} ;
		\draw[ultra thick] (1,0)  node[cross=5pt,rotate=0] {} ;
		\draw[ultra thick] (2,-1)  node[cross=5pt,rotate=-45] {} ;
		\filldraw[ultra thick] (4,0)circle (2pt) node[align=center, below] {$\phi_{(0)}(\omega)$} -- (3,0)circle (2pt) node[align=left, below] {};
		\filldraw[ultra thick] (2,0)circle (2pt) node[align=right, above]{} -- (1,1)circle (0pt) node[align=center, above] {} ; 
		\filldraw[ultra thick] (2,0) -- (1,-1)circle (0pt) node[align=center, below] {};
		\filldraw[ultra thick] (3,0) -- (1,0)circle (0pt) node[align=center, below] {};
		
		\filldraw[ultra thick] (3,0) -- (2,1)circle (0pt) node[align=center, above] {};
		\filldraw[ultra thick] (3,0) -- (2,-1)circle (0pt) node[align=center, below] {}; 
		\end{tikzpicture}  \\ 
		\hline
	\end{tabular}
\end{center}

The 15 Wick different ways into which  $\left\langle\tilde{\xi}(\omega_1)\cdot\tilde{\xi}(\omega_2)\cdot\xi(\omega_3)\cdot\xi(\omega_1')\cdot\xi(\omega_2')\cdot\xi(\omega_3')\right\rangle$ can be expanded out split \eqref{phi2phi0} into 3 topologically different diagrams:
3 "Right Double Seagull", 6 "Right Sunset" and 6 "Right Cactus" diagrams (and the corresponding "Left" ones from the $\omega \rightarrow \omega'$ symmetric.)

\textbf{The Right Double Seaguls}\\
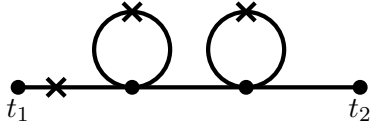
\begin{minipage}{5cm}
	\begin{tikzpicture}
	\filldraw[ultra thick](5,0)circle (2pt) node[align=center, below] {$t_2$} ;
	\filldraw[ultra thick](3.5,0)circle (2pt) node[align=center, below] {$$} ;
	\filldraw[ultra thick](2,0)circle (2pt) node[align=center, below] {$$} ;
	\filldraw[ultra thick](0.5,0)circle (2pt) node[align=center, below] {$t_1$} ;
	\filldraw [ultra thick](0.5,0) -- (5,0);
	\draw [ultra thick] (2,0.5) circle [radius=0.5];
	\draw [ultra thick] (3.5,0.5) circle [radius=0.5];
	\draw[ultra thick] (2,1)  node[cross=5pt,rotate=0]{} ;
	\draw[ultra thick] (3.5,1)  node[cross=5pt,rotate=0]{} ;
	\draw[ultra thick] (1,0)  node[cross=5pt,rotate=0]{} ;
	\end{tikzpicture}
\end{minipage}
\begin{minipage}{10cm}
	\begin{equation}
	\begin{split}
	\mathcal{RDS}&=3 \cdot \frac{ 2 \lambda^2 \, \hbar H_0^7}{27\cdot \left(8\pi^2\right)^3} G^R(\omega) G^R(\omega') \int \frac{d\omega_1 d\omega_2 d\omega_3 d\omega'_1 d\omega'_2 d\omega'_3 }{(2\pi)^6}\\
	&\hspace{0.5cm}\,\slashed{\delta}(\omega_1+\omega_2+\omega_3-\omega_2') \slashed{\delta}(\omega_1'+\omega_2'+\omega_3'-\omega')\\
	&\hspace{0.5cm}\cdot  G^R(\omega_1) \,G^R(\omega_2) \,G^R(\omega_3) \,G^R(\omega_1') \,G^R(\omega_2') \,G^R(\omega_3')\\
	&\hspace{0.5cm}\cdot \slashed{\delta}(\omega_1'+\omega_3')\slashed{\delta}(\omega_1+\omega_3)\slashed{\delta}(\omega_2+\omega)\\
	&=\frac14 \, \, \, \frac{2 \, \lambda^2 H_0^7 \,\hbar }{9\slashed{m}^2 \,\, \, \left(8\pi^2\right)^3 }  \int \frac{d\omega}{2\pi} \frac{e^{i \omega\left(t_2-t_1\right)}}{\left(\slashed{m}^2+\omega^2\right) \left(\slashed{m}-i\omega\right) \ \left(\slashed{m}-i\omega\right)}
	\end{split}
	\end{equation}
\end{minipage}

\textbf{The Right Sunset}\\

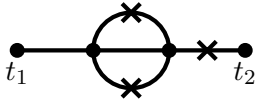
\begin{minipage}{5cm}
	\begin{tikzpicture}
	\filldraw[ultra thick](3,0)circle (2pt) node[align=center, below] {$t_2$} ;
	\filldraw[ultra thick](2,0)circle (2pt) node[align=center, below] {$\hspace{0.3cm}$} ;
	\filldraw[ultra thick](1,0)circle (2pt) node[align=center, below] {$\hspace{-0.3cm}$} ;
	\filldraw[ultra thick](0,0)circle (2pt) node[align=center, below] {$t_1$} ;
	\filldraw [ultra thick](0,0) -- (3,0);
	\draw [ultra thick] (1.5,0) circle [radius=0.5];
	\draw[ultra thick] (1.5,0.5)  node[cross=5pt,rotate=0]{} ;
	\draw[ultra thick] (1.5,-0.5)  node[cross=5pt,rotate=0]{} ;
	\draw[ultra thick] (2.5,0)  node[cross=5pt,rotate=0]{} ;
	\end{tikzpicture}
\end{minipage}
\begin{minipage}{10cm}
	\begin{equation}
	\begin{split}
	\mathcal{RS}&= 6\cdot \frac{ 2 \lambda^2 \, \hbar H_0^7}{27\cdot \left(8\pi^2\right)^3} G^R(\omega) G^R(\omega') \int \frac{d\omega_1 d\omega_2 d\omega_3 d\omega'_1 d\omega'_2 d\omega'_3 }{(2\pi)^6}\\
	&\hspace{0.5cm}\,\slashed{\delta}(\omega_1+\omega_2+\omega_3-\omega_2') \slashed{\delta}(\omega_1'+\omega_2'+\omega_3'-\omega')\\
	&\hspace{0.5cm}\cdot  G^R(\omega_1) \,G^R(\omega_2) \,G^R(\omega_3) \,G^R(\omega_1') \,G^R(\omega_2') \,G^R(\omega_3')\\
	&\hspace{0.5cm}\cdot \slashed{\delta}(\omega_1'+\omega_3')\slashed{\delta}(\omega_1+\omega_3)\slashed{\delta}(\omega_2+\omega)\\
	&=\frac{ \lambda^2 H_0^7 \,\hbar }{9\slashed{m}^2 \,\, \, \left(8\pi^2\right)^3 } \int \frac{d\omega}{2\pi} \frac{e^{i \omega\left(t_2-t_1\right)}  \ (\slashed{m}+i\omega)(3\slashed{m}+i\omega)  }{\left(\slashed{m}^2+\omega^2\right) \left[ \left(3\slashed{m}\right)^2 +\omega^2\right] }
	\end{split}
	\end{equation}
\end{minipage}

\textbf{The Right Cactus}\\

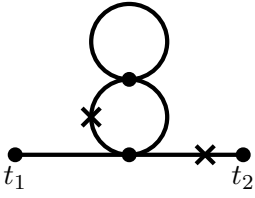
\begin{minipage}{5cm}
	\begin{tikzpicture}
	\filldraw[ultra thick](-1.5,0)circle(2pt) node[align=center, below] {$t_1$};
	\filldraw[ultra thick](0,0)circle (2pt) node[align=center, below] {$$};
	\filldraw[ultra thick](1.5,0)circle (2pt) node[align=center, below] {$t_2$};
	\filldraw[ultra thick](0,1)circle (2pt) node[align=center, below] {$$};
	\filldraw[ultra thick] (-1.5,0)-- (0,0);
	\filldraw[ultra thick] (0,0)-- (1.5,0);
	\draw [ultra thick] (0,1.5) circle [radius=0.5];
	\draw [ultra thick] (0,0.5) circle[radius=0.5];	
	\draw[ultra thick] (1,0)  node[cross=5pt,rotate=0]{} ;
	\draw[ultra thick] (-0.5,0.5)  node[cross=5pt,rotate=0] {} ;
	\end{tikzpicture}
\end{minipage}
\begin{minipage}{12cm}
	\begin{equation}
	\begin{split}
	\mathcal{RC}&=6 \cdot \frac{ 2 \lambda^2 \, \hbar H_0^7}{27\cdot \left(8\pi^2\right)^3} G^R(\omega) G^R(\omega') \int \frac{d\omega_1 d\omega_2 d\omega_3 d\omega'_1 d\omega'_2 d\omega'_3 }{(2\pi)^6}\\
	&\hspace{0.5cm}\,\slashed{\delta}(\omega_1+\omega_2+\omega_3-\omega_2') \slashed{\delta}(\omega_1'+\omega_2'+\omega_3'-\omega')\\
	&\hspace{0.5cm}\cdot  G^R(\omega_1) \,G^R(\omega_2) \,G^R(\omega_3) \,G^R(\omega_1') \,G^R(\omega_2') \,G^R(\omega_3')\\
	&\hspace{0.5cm}\cdot \slashed{\delta}(\omega+\omega_3')\slashed{\delta}(\omega_2+\omega_3)\slashed{\delta}(\omega_1'+\omega_1)\\
	&\hspace{0.5cm}=\frac{ \lambda^2 H_0^7 \, \hbar }{18\slashed{m}^3 \,\, \, \left(8\pi^2\right)^3 }   \int \frac{d\omega}{2\pi} \frac{e^{i \omega\left(t_2-t_1\right)} \, \slashed{m}-i\omega }{\left(\slashed{m}^2+\omega^2\right)^2},
	\end{split}
	\end{equation}
\end{minipage}

All are of course in direct one-to-one agreement with their path-integral counterparts \eqref{rightsseagull},\eqref{rightsun} and \eqref{rightcactus}.

It is obvious to see, taking $\omega \rightarrow -\omega$ (and changing the direction of time) that the time-symmetric
diagrams \eqref{leftsseagull}, \eqref{leftsun} and \eqref{leftcactus} are obtained, again in one-to-one agreement.

\subsection{O($\lambda^2$) Four-Point Function}

The first order ($O^{\lambda^2}$) correction to the four-field vertex can be calculated directly, from the solutions of the stochastic differential equations. Then, the four point function can be expanded as follows:

\begin{equation}
\begin{split}
\label{4point}
\left\langle\phi(t_1)\phi(t_2)\phi(t_3)\phi(t_4)\right\rangle^{O(\lambda^2)}
=&\,\, \, \, \left\langle\phi_0(t_1)\phi_0(t_2)\phi_1(t_3)\phi_1(t_4)\right\rangle
+\left\langle\phi_1(t_1)\phi_1(t_2)\phi_0(t_3)\phi_0(t_4)\right\rangle\\
&+\left\langle\phi_0(t_1)\phi_1(t_2)\phi_1(t_3)\phi_0(t_4)\right\rangle
+\left\langle\phi_1(t_1)\phi_0(t_2)\phi_0(t_3)\phi_1(t_4)\right\rangle\\
&+\left\langle\phi_1(t_1)\phi_0(t_2)\phi_1(t_3)\phi_0(t_4)\right\rangle
+\left\langle\phi_0(t_1)\phi_1(t_2)\phi_0(t_3)\phi_1(t_4)\right\rangle\\
&+\left\langle\phi_2(t_1)\phi_0(t_2)\phi_0(t_3)\phi_0(t_4)\right\rangle
+\left\langle\phi_0(t_1)\phi_2(t_2)\phi_0(t_3)\phi_0(t_4)\right\rangle\\
&+\left\langle\phi_0(t_1)\phi_0(t_2)\phi_2(t_3)\phi_0(t_4)\right\rangle
+\left\langle\phi_0(t_1)\phi_0(t_2)\phi_0(t_3)\phi_2(t_4)\right\rangle.
\end{split}
\end{equation}
We group the terms in the first, second and third lines as well as the four last terms as leading to topologically different types of "candy" diagram and show that, as expected, the final results are identical to those presented in section \eqref{picandies}.

\subsubsection{F-Candies}
The diagrams containing an F-loop can be seperated depending on their topology in "Horizontal", "Vertical" and "Knotted" Candies. We explicitly show the calculation of the first type and present the result of the computation of the rest. 

In order to calculate the first term of \eqref{4point}, $\left\langle\phi_0(t_1)\phi_0(t_2)\phi_1(t_3)\phi_1(t_4)\right\rangle$, we anchor the incoming particles as states $^{1}_3$ and the outgoing as $^2_4$ :

\begin{equation}
\begin{split}
\left\langle\phi_0(t_1)\phi_0(t_2)\phi_1(t_3)\phi_1(t_4)\right\rangle=& 18 \int\frac{d\omega_1 \,d\omega_2 \,d\omega_3 \,d\omega_4 \,d\omega'_1\,d\omega'_2\,d\omega'_3\,d\tilde{\omega}_1\,d\tilde{\omega}_2\,d\tilde{\omega}_3 \, \,}{(2\pi)^{10} } \\
&\cdot \left( \frac{\lambda}{18\,H\, \hbar }   \right)^2 \,\, \left(\frac{H_0^3 \, \hbar}{4\pi^2}\right)^4 \, e^{i(\omega_i t^i)} \slashed{\delta}(\omega_1+\omega'_1)\, \slashed{\delta}(\tilde{\omega}_2+\omega'_2)\\
&\cdot\slashed{\delta}(\tilde{\omega}_3+\omega'_3)  \, \slashed{\delta}(\tilde{\omega}_1+\omega_2)  \,  \slashed{\delta}(\omega'_1+\omega'_2+\omega'_3-\omega_3)\,\,\\
&\cdot\slashed{\delta}(\tilde{\omega}_1+\tilde{\omega}_2+\tilde{\omega}_3-\omega_4) 
G^R(\omega_1)G^R(\omega_2)G^R(\omega_3)G^R(\omega_4)\\
&\cdot G^R(\omega'_1)G^R(\omega'_2)G^R(\omega'_3)G^R(\slashed{\omega}_1)G^R(\slashed{\omega}_2)G^R(\slashed{\omega}_3)\\
=&18 \left( \frac{\lambda}{18\,H\, \hbar }   \right)^2 \,\, \left(\frac{H_0^3 \, \hbar }{4\pi^2}\right)^4 \int \frac{d\omega_1 \,d\omega_2 \,d\omega_3 \,d\omega_4}{(2\pi)^4} \, \frac{e^{i(\sum\omega_i t_i)}}{(2\slashed{m})^2 +(\omega_1+\omega_3)^2}\\
&\cdot G^R(\omega_1)G^R(-\omega_1)G^R(\omega_2)G^R(-\omega_2) G^R(\omega_3)G^R(\omega_4)\slashed{\delta}(\Sigma \omega_i)
\end{split}
\end{equation}
The origin of the factor of 18 comes from the topologically equivalent Wick contractions:
There are 3 distinct choices for $t_1$ to be linked to any of the three prongs of $t_3$ and the same holds for $t_2$ and $t_4$ (resulting in a factor of 9). The two left over prongs of each of $t_3$ and $t_4$ can form a loop in two different ways (resulting in a multiplicative factor of 2). Hence, there are 18 different wick contractions of the eight $\xi$s that preserve the structure of the external times as described earlier.
The last line reproduces the path integral result \eqref{C1} exactly.

Performing the following permutations $(1\leftrightarrow3,2\leftrightarrow4)$, ($1\leftrightarrow3$), ($2\leftrightarrow4$) and adding up the individual diagram contributions, results in:

\begin{equation}
\begin{split}
&\left\langle\phi_0(t_1)\phi_0(t_2)\phi_1(t_3)\phi_1(t_4)\right\rangle +	\left\langle\phi_1(t_1)\phi_1(t_2)\phi_0(t_3)\phi_0(t_4)\right\rangle\\
&\left\langle\phi_1(t_1)\phi_0(t_2)\phi_0(t_3)\phi_1(t_4)\right\rangle +	\left\langle\phi_0(t_1)\phi_1(t_2)\phi_1(t_3)\phi_0(t_4)\right\rangle\\
&=\frac{1}{2\slashed{m}} \left[\frac{\lambda}{3\hbar \, H_0 }\right]^2 
\int \frac{d\omega_1 \, d\omega_2 \, d \omega_3 \, d \omega_4}{(2\pi)^4} \prod_i F(\omega_i)\, e^{i \Sigma \left(\omega_i t_i\right)} \slashed{\delta}\left(\Sigma \omega_i\right)
\end{split}
\end{equation}

In exact agreement with the Path Integral method result, \eqref{horsum}.
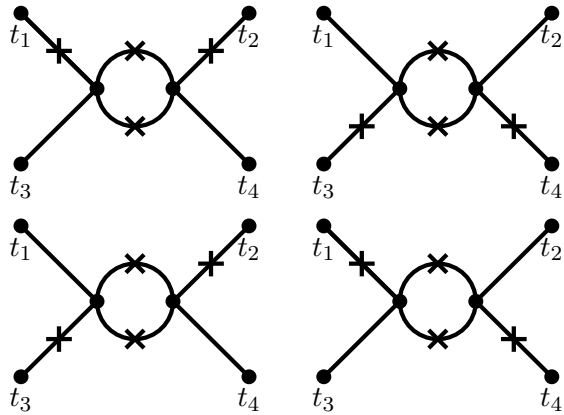
\begin{table}[h]
	\begin{center}
		\begin{tabular}{ c c} 
			\begin{tikzpicture}
			\filldraw[ultra thick] (0,0)circle (2pt) node[align=center, right]{$$};
			\filldraw[ultra thick] (1,0)circle (2pt) node[align=center, left]{$$}; 
			\filldraw[ultra thick] (-1,1)circle (2pt) node[align=center, below]{$t_1$};
			\filldraw[ultra thick] (-1,-1)circle (2pt) node[align=center, below]{$t_3$};
			\filldraw[ultra thick] (2,1)circle (2pt) node[align=center, below]{$t_2$};
			\filldraw[ultra thick] (2,-1)circle (2pt) node[align=center, below]{$t_4$};
			\filldraw[ultra thick](-1,1)--(0,0);
			\filldraw[ultra thick](-1,-1)--(0,0);
			\filldraw[ultra thick](2,1)--(1,0);
			\filldraw[ultra thick](2,-1)--(1,0);
			\filldraw[ultra thick](-1,1)--(0,0);
			\draw [ ultra thick] (0.5,0) circle(0.5);
			\draw[ultra thick] (-0.5,0.5)  node[cross=5pt,rotate=45] {} ;
			\draw[ultra thick] (1.5 , 0.5)  node[cross=5pt,rotate=45] {} ;
			\draw[ultra thick] (0.5,0.5)  node[cross=5pt,rotate=0] {} ;
			\draw[ultra thick] (0.5,-0.5)  node[cross=5pt,rotate=0] {} ;
			\end{tikzpicture}	
			&
			\begin{tikzpicture}
			\filldraw[ultra thick] (0,0)circle (2pt) node[align=center, right]{$$};
			\filldraw[ultra thick] (1,0)circle (2pt) node[align=center, left]{$$}; 
			\filldraw[ultra thick] (-1,1)circle (2pt) node[align=center, below]{$t_1$};
			\filldraw[ultra thick] (-1,-1)circle (2pt) node[align=center, below]{$t_3$};
			\filldraw[ultra thick] (2,1)circle (2pt) node[align=center, below]{$t_2$};
			\filldraw[ultra thick] (2,-1)circle (2pt) node[align=center, below]{$t_4$};
			\filldraw[ultra thick](1,0)--(2,-1);
			\filldraw[ultra thick](-1,-1)--(0,0);
			\filldraw[ultra thick](2,1)--(1,0);
			\filldraw[ultra thick](-1,1)--(0,0);
			\draw [ ultra thick] (0.5,0) circle(0.5);
			\draw[ultra thick] (-0.5,-0.5)  node[cross=5pt,rotate=45] {} ;
			\draw[ultra thick] (1.5,-0.5)  node[cross=5pt,rotate=45] {} ;
			\draw[ultra thick] (0.5,0.5)  node[cross=5pt,rotate=0] {} ;
			\draw[ultra thick] (0.5,-0.5)  node[cross=5pt,rotate=0] {} ;
			\end{tikzpicture}
			\\	
			\begin{tikzpicture}
			\filldraw[ultra thick] (0,0)circle (2pt) node[align=center, right]{$$};
			\filldraw[ultra thick] (1,0)circle (2pt) node[align=center, left]{$$}; 
			\filldraw[ultra thick] (-1,1)circle (2pt) node[align=center, below]{$t_1$};
			\filldraw[ultra thick] (-1,-1)circle (2pt) node[align=center, below]{$t_3$};
			\filldraw[ultra thick] (2,1)circle (2pt) node[align=center, below]{$t_2$};
			\filldraw[ultra thick] (2,-1)circle (2pt) node[align=center, below]{$t_4$};
			\filldraw[ultra thick](-1,1)--(0,0);
			\filldraw[ultra thick](-1,-1)--(0,0);
			\filldraw[ultra thick](2,1)--(1,0);
			\filldraw[ultra thick](2,-1)--(1,0);
			\draw [ ultra thick] (.5,0) circle(0.5);
			\draw[ultra thick] (-0.5,-0.5)  node[cross=5pt,rotate=45] {} ;
			\draw[ultra thick] (1.5,0.5)  node[cross=5pt,rotate=45] {} ;
			\draw[ultra thick] (0.5,0.5)  node[cross=5pt,rotate=0] {} ;
			\draw[ultra thick] (0.5,-0.5)  node[cross=5pt,rotate=0] {} ;
			\end{tikzpicture}		
			&
			\begin{tikzpicture}
			\filldraw[ultra thick] (0,0)circle (2pt) node[align=center, right]{$$};
			\filldraw[ultra thick] (1,0)circle (2pt) node[align=center, left]{$$}; 
			\filldraw[ultra thick] (-1,1)circle (2pt) node[align=center, below]{$t_1$};
			\filldraw[ultra thick] (-1,-1)circle (2pt) node[align=center, below]{$t_3$};
			\filldraw[ultra thick] (2,1)circle (2pt) node[align=center, below]{$t_2$};
			\filldraw[ultra thick] (2,-1)circle (2pt) node[align=center, below]{$t_4$};
			\filldraw[ultra thick](-1,1)--(0,0);
			\filldraw[ultra thick](-1,-1)--(0,0);
			\filldraw[ultra thick](2,1)--(1,0);
			\filldraw[ultra thick](2,-1)--(1,0);
			\filldraw[ultra thick](-1,1)--(0,0);
			\draw [ ultra thick] (0.5,0) circle(0.5);
			\draw[ultra thick] (-0.5,0.5)  node[cross=5pt,rotate=45] {} ;
			\draw[ultra thick] (1.5,-0.5)  node[cross=5pt,rotate=45] {} ;
			\draw[ultra thick] (0.5,0.5)  node[cross=5pt,rotate=0] {} ;
			\draw[ultra thick] (0.5,-0.5)  node[cross=5pt,rotate=0] {} ;
			\end{tikzpicture}
			\\
		\end{tabular}
	\end{center}
\end{table}

Furthermore, choosing to connect one of the prongs of the $\phi_{(1)}(t_1)$ with $\phi_0(t_2)$ forms the "vertical candy" diagram.
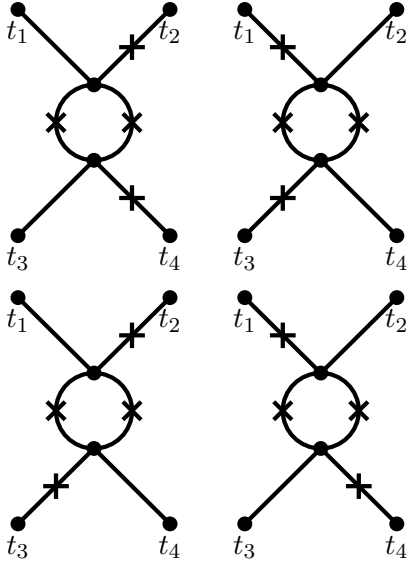
\begin{table}[h]
	\begin{center}
		\begin{tabular}{cc} 
			\begin{tikzpicture}
			\filldraw[ultra thick] (0,1)circle (2pt) node[align=center, below]{$$};
			\filldraw[ultra thick] (0,0)circle (2pt) node[align=center, above]{$$}; 
			\filldraw[ultra thick] (-1,2)circle (2pt) node[align=center, below]{$t_1$};
			\filldraw[ultra thick] (1,2)circle (2pt) node[align=center, below]{$t_2$};
			\filldraw[ultra thick] (-1,-1)circle (2pt) node[align=center, below]{$t_3$};
			\filldraw[ultra thick] (1,-1)circle (2pt) node[align=center, below]{$t_4$};
			\filldraw[ultra thick](0,1)--(-1,2);
			\filldraw[ultra thick](1,2)--(0,1);
			\filldraw[ultra thick](0,0)--(-1,-1);
			\filldraw[ultra thick](1,-1)--(0,0);
			\draw [ ultra thick] (0,0.5) circle(0.5);	
			\draw[ultra thick] (0.5,-0.5)  node[cross=5pt,rotate=45] {} ;
			\draw[ultra thick] (0.5,1.5)  node[cross=5pt,rotate=45] {} ;
			\draw[ultra thick] (0.5,0.5)  node[cross=5pt,rotate=0] {} ;
			\draw[ultra thick] (-0.5,0.5)  node[cross=5pt,rotate=0] {} ;
			\end{tikzpicture}	
			&
			\begin{tikzpicture}
			\filldraw[ultra thick] (0,1)circle (2pt) node[align=center, below]{$$};
			\filldraw[ultra thick] (0,0)circle (2pt) node[align=center, above]{$$}; 
			\filldraw[ultra thick] (-1,2)circle (2pt) node[align=center, below]{$t_1$};
			\filldraw[ultra thick] (1,2)circle (2pt) node[align=center, below]{$t_2$};
			\filldraw[ultra thick] (-1,-1)circle (2pt) node[align=center, below]{$t_3$};
			\filldraw[ultra thick] (1,-1)circle (2pt) node[align=center, below]{$t_4$};
			\filldraw[ultra thick](0,1)--(-1,2);
			\filldraw[ultra thick](1,2)--(0,1);
			\filldraw[ultra thick](0,0)--(-1,-1);
			\filldraw[ultra thick](1,-1)--(0,0);
			\draw [ ultra thick] (0,0.5) circle(0.5);	
			\draw[ultra thick] (-0.5,1.5)  node[cross=5pt,rotate=45] {} ;
			\draw[ultra thick] (-0.5,-0.5)  node[cross=5pt,rotate=45] {} ;
			\draw[ultra thick] (0.5,0.5)  node[cross=5pt,rotate=0] {} ;
			\draw[ultra thick] (-0.5,0.5)  node[cross=5pt,rotate=0] {} ;
			\end{tikzpicture}\\	
			\begin{tikzpicture}
			\filldraw[ultra thick] (0,1)circle (2pt) node[align=center, below]{$$};
			\filldraw[ultra thick] (0,0)circle (2pt) node[align=center, above]{$$}; 
			\filldraw[ultra thick] (-1,2)circle (2pt) node[align=center, below]{$t_1$};
			\filldraw[ultra thick] (1,2)circle (2pt) node[align=center, below]{$t_2$};
			\filldraw[ultra thick] (-1,-1)circle (2pt) node[align=center, below]{$t_3$};
			\filldraw[ultra thick] (1,-1)circle (2pt) node[align=center, below]{$t_4$};
			\filldraw[ultra thick](0,1)--(-1,2);
			\filldraw[ultra thick](1,2)--(0,1);
			\filldraw[ultra thick](0,0)--(-1,-1);
			\filldraw[ultra thick](1,-1)--(0,0);
			\draw [ ultra thick] (0,0.5) circle(0.5);	
			\draw[ultra thick] (0.5,1.5)  node[cross=5pt,rotate=45] {} ;
			\draw[ultra thick] (-0.5,-0.5)  node[cross=5pt,rotate=45] {} ;
			\draw[ultra thick] (0.5,0.5)  node[cross=5pt,rotate=0] {} ;
			\draw[ultra thick] (-0.5,0.5)  node[cross=5pt,rotate=0] {} ;
			\end{tikzpicture}
			&
			
			\begin{tikzpicture}
			\filldraw[ultra thick] (0,1)circle (2pt) node[align=center, below]{$$};
			\filldraw[ultra thick] (0,0)circle (2pt) node[align=center, above]{$$}; 
			\filldraw[ultra thick] (-1,2)circle (2pt) node[align=center, below]{$t_1$};
			\filldraw[ultra thick] (1,2)circle (2pt) node[align=center, below]{$t_2$};
			\filldraw[ultra thick] (-1,-1)circle (2pt) node[align=center, below]{$t_3$};
			\filldraw[ultra thick] (1,-1)circle (2pt) node[align=center, below]{$t_4$};
			\filldraw[ultra thick](0,1)--(-1,2);
			\filldraw[ultra thick](1,2)--(0,1);
			\filldraw[ultra thick](0,0)--(-1,-1);
			\filldraw[ultra thick](1,-1)--(0,0);
			\draw [ ultra thick] (0,0.5) circle(0.5);	
			\draw[ultra thick] (-0.5,1.5)  node[cross=5pt,rotate=45] {} ;
			\draw[ultra thick] (0.5,-0.5)  node[cross=5pt,rotate=45] {} ;
			\draw[ultra thick] (0.5,0.5)  node[cross=5pt,rotate=0] {} ;
			\draw[ultra thick] (-0.5,0.5)  node[cross=5pt,rotate=0] {} ;
			\end{tikzpicture}
		\end{tabular}
	\end{center}
\end{table}

Similarly to the "Horizontal Candy" diagrams, the sum of the four vertical ones results in:

\begin{equation}
\sum_i \mathbf{\mathcal{V}}_i=\frac{1}{2\slashed{m}} \left[\frac{\lambda}{3\hbar \, H_0 }\right]^2 
\int \frac{d\omega_1 \, d\omega_2 \, d \omega_3 \, d \omega_4}{(2\pi)^4} \prod_i F(\omega_i)\, e^{i \Sigma \left(\omega_i t_i\right)} \slashed{\delta}\left(\Sigma \omega_i\right),
\end{equation}

Lastly, the knotted candies give 
\begin{equation}
\sum_i K_i=\frac{1}{2\slashed{m}} \left[\frac{\lambda}{3\hbar \, H_0 }\right]^2 
\int \frac{d\omega_1 \, d\omega_2 \, d \omega_3 \, d \omega_4}{(2\pi)^4} \prod_i F(\omega_i)\, e^{i \Sigma \left(\omega_i t_i\right)} \slashed{\delta}\left(\Sigma \omega_i\right)
\end{equation}

Ultimately, adding up all the contributions we obtain for the sum of the F-Candies: 

\begin{equation}
\sum_i \mathcal{F}_i=\frac{3}{2\slashed{m}} \left[\frac{\lambda}{3\hbar \, H_0 }\right]^2 
\int \frac{d\omega_1 \, d\omega_2 \, d \omega_3 \, d \omega_4}{(2\pi)^4} \prod_i F(\omega_i)\, e^{i \Sigma \left(\omega_i t_i\right)} \slashed{\delta}\left(\Sigma \omega_i\right)
\end{equation}

\subsubsection{Loop $G$ Candies}
The four last contributions in \eqref{4point} form a different type of correction to the four-point vertex, namely one in which one of the one set of the three prongs of the $\phi_2(t)$ closes in with one of the other set of 3 prongs. This will lead to diagrams in which one of the internal loop propagators is $F(\omega)$ and one that is $G$ in the language of section \ref{sect:path integral}.

Starting with $\mathcal{A}=<\phi_2(t_1)\phi_0(t_2)\phi_0(t_3)\phi_0(t_4)>$

\begin{equation}
\begin{split}
\mathcal{A}=&36 \,\cdot \,  3\left( \frac{\lambda}{18\,H\, \hbar }   \right)^2 \,\, \left(\frac{H_0^3 \, \hbar}{4\pi^2}\right)^4 \int \frac{d\omega_1 \, d\omega_2 \, d\omega_3 \, d\omega_4 \, d\omega'_1 \, d\omega'_2 \, d\omega'_3\, d\tilde{\omega}_1\, d\tilde{\omega}_2\, d\tilde{\omega}_3     }{(2\pi)^{10}} \, \,  e^{i(\Sigma \omega_i\, t_i)}  \, \\ 
&\cdot   \slashed{\delta}(\omega'_1+\tilde{\omega}_1) \, \slashed{\delta}(\omega_2+\tilde{\omega}_2) \, \slashed{\delta}(\omega_4+\tilde{\omega}_3) \,\slashed{\delta}(\omega'_2+{\omega}_3) \\
&\cdot\slashed{\delta}(\tilde{\omega}_1+\tilde{\omega}_2+\tilde{\omega}_3-\omega'_3)
\slashed{\delta}(\omega'_1+\omega'_2+\omega'_3-\omega_1)  \\
&\cdot G^R(\omega_1) \, G^R(\omega_2) \, G^R(\omega_3) \, G^R(\omega_4) \, G^R(\omega'_1) \, G^R(\omega'_2) \, G^R(\omega'_3) \, G^R(\tilde{\omega}_1)G^R(\tilde{\omega}_2)G^R(\tilde{\omega}_3) \\
=&36 \,\cdot \,  3\left( \frac{\lambda}{18\,H\, \hbar }   \right)^2 \,\, \left(\frac{H_0^3 \, \hbar}{4\pi^2}\right)^4  \int \frac{d\omega_1 \, d\omega_2 \, d\omega_3 \, d\omega_4 }{(2\pi)^{4}} \,  e^{i(\Sigma \omega_i\, t_i)} \frac{1}{2\slashed{m}} \frac{\slashed{m}-i\omega_1}{(\slashed{2m})+i(\omega_2+\omega_4)} \\
&\cdot  \, G^R(\omega_2)\, G^R(-\omega_2)\, G^R(\omega_3)\, G^R(-\omega_3)\, G^R(\omega_4)\, G^R(-\omega_4) \slashed{\delta}(\Sigma \omega_i)
\end{split}
\end{equation}
The other 3 diagrams are obtained by performing the following permutations: 

\begin{equation}\begin{split}
\mathcal{B}&=\mathcal{A}, (1\leftrightarrow 3)\\
\mathcal{C}&=\mathcal{A}, (1\leftrightarrow 2, 3\leftrightarrow 4)\\
\mathcal{D}&=\mathcal{A}, (1\leftrightarrow  4, 2\leftrightarrow 3)\\
\end{split}
\end{equation}
Then, the sum of the 4 diagrams, is: 

\begin{equation}
\begin{split}
\mathcal{F}=&\mathcal{A}+\mathcal{B}+\mathcal{C}+\mathcal{D}\\
=&36 \,\cdot \,  \frac{3}{2\slashed{m}}\left( \frac{\lambda}{18\,H\, \hbar }   \right)^2 \,\, \left(\frac{H_0^3 \, \hbar}{4\pi^2}\right)^4  \int \frac{d\omega_1 \, d\omega_2 \, d\omega_3 \, d\omega_4 }{(2\pi)^{4}} \,  e^{i(\Sigma \omega_i\, t_i)} F(\omega_1) \, F(\omega_2)\,  F(\omega_3)\,F(\omega_4) \slashed{\delta}(\Sigma \omega_i)\\
&\cdot \left[ \frac{\slashed{m}-i\omega_1}{(\slashed{2m})+i(\omega_2+\omega_4)} + \frac{\slashed{m}-i\omega_3}{(\slashed{2m})+i(\omega_2+\omega_4)} + \frac{\slashed{m}-i\omega_2}{(\slashed{2m})+i(\omega_1+\omega_3)} + \frac{\slashed{m}-i\omega_4}{(\slashed{2m})+i(\omega_1+\omega_3)} \right] \\
=& \, \frac{3}{\slashed{m}} \, \left[\frac{\lambda}{3\hbar \, H_0 }\right]^2   \int \frac{d\omega_1 \, d\omega_2 \, d\omega_3 \, d\omega_4 }{(2\pi)^{4}} \,  e^{i(\Sigma \omega_i\, t_i)} \prod_i F(\omega_i) \slashed{\delta}(\Sigma \omega_i), 
\end{split}
\end{equation}
as expected from \eqref{grsum}.

Here, the multiplicative factor of 36 is due to the Wick contractions as follows: 
There are 3 $\tilde{\omega_i}$ prongs that can form the $G^R$-loop with one of the two $\omega'_i$ prongs, giving rise to a factor of 6. The two left over of the $\tilde{\omega_i}$ are interchangable (resulting in a factor of 2) and lastly, there is a single way to connect the remaining $\omega'_i$ prong with any one of $\omega_2$,$\omega_3$ or $\omega_4$, giving a horizontal, vertical or knotted G-candy, respectively (another factor of 3).
Hence, the total one-loop correction to the 4-point vertex is:

\begin{equation}
\mathcal{Z}=\frac92 \left(\frac{\lambda}{3\hbar H}\right)^2 \, \frac{H^{12}}{(4\pi^2)^4\slashed{m}}\,  \int \frac{d\omega_1 \, d\omega_2 \, d \omega_3 \, d \omega_4}{(2\pi)^4} \,\cdot \slashed{\delta}(\Sigma \omega_i) \, e^{i(\Sigma \omega_i\, t_i)} \, \prod F(\omega_i)
\end{equation}
as previously found.

\section{Backreaction contributions to the Two-Point Correlators}\label{Gravitational contributions}

In this section we compute corrections stemming from the $\phi$ dependence in the noise amplitude $\mathcal{A}\propto H^3$, making it multiplicative type noise. We note that we neglect any new vertices that are further suppressed by $\frac{m^2}{H_0^2}$ or $\lambda$. Furthermore, we have not computed corrections to the noise amplitude due to the modified behaviour of the scalar modes as they exit the horizon which are presumably suppressed by similar factors. This computation therefore serves as an illustration of the new types of vertices that arise due to the gravitational backreaction of the field $\phi$, but should also contain the leading order result. The new contributions are easy to compute in the path integral formalism where a new set of vertices appears, see section \ref{sect:path integral}. These contribute to the two point function to leading order as follows:

\vspace{1cm}
\begin{minipage}[b]{0.45\textwidth}
	\hspace{-1cm}
	\begin{tabular}{ cc } 	
		\begin{tikzpicture}
		\filldraw[ultra thick] (-1.5,0)circle (2pt) node[align=center, below] {$t_1$} -- (-1,0);
		\draw[ultra thick,snake] (-1,0) -- (0,0)circle (2pt);
		circle (2pt)
		\filldraw [  ultra thick  ] (0,0)circle (2pt) node[align=center, below]{$$} -- (1.5,0)circle (2pt) node[align=center, below] {$t_2$};
		\draw [ dashed, ultra thick] (0,0.5) circle [radius=0.5];
		\draw[->, ultra thick, dashed] (0,1)--(-0.01,1);
		\end{tikzpicture}
		&
		\begin{tikzpicture}
		\filldraw[ultra thick] (-1.5,0)circle (2pt) node[align=center, below] {$t_1$} -- (-1,0);
		\filldraw[ultra thick] (-1,0) -- (0,0)circle (2pt);
		circle (2pt)
		\draw [  ultra thick,snake  ] (0,0)circle (2pt) node[align=center, below]{$$} -- (1,0);
		\filldraw [  ultra thick ](1,0)-- (1.5,0)circle (2pt) node[align=center, below] {$t_2$};
		\draw [ dashed, ultra thick] (0,0.5) circle [radius=0.5];
		\draw[->, ultra thick, dashed] (0,1)--(-0.01,1);
		\end{tikzpicture}    \\ 
		\begin{tikzpicture}
		\filldraw[ultra thick] (-1.5,0)circle (2pt) node[align=center, below] {$t_1$} -- (-1,0);
		\draw [  ultra thick,snake ] (-1,0)-- (0,0)circle (2pt) node[align=center, below]{$$} ;
		\filldraw [  ultra thick  ] (0,0) -- (1.5,0)circle (2pt) node[align=center, below] {$t_2$};
		\draw [ ultra thick] (0,1) arc (90:270:0.5cm);
		\draw [get length](0,0) arc (270:90:0.5cm) ;
		\draw[ultra thick,decoration={snake,amplitude=0.06cm,segment length={\pathlen /7}},
		decorate] (0,0) arc (-90:90:0.5cm);
		\end{tikzpicture}
		&
		\begin{tikzpicture}
		\filldraw[ultra thick ] (-1.5,0)circle (2pt) node[align=center, below] {$t_1$} -- (0,0)circle (2pt) node[align=center, below]{$$};
		\draw[ultra thick,snake] (0,0)-- (1,0);
		\filldraw [  ultra thick ] (1,0) -- (1.5,0)circle (2pt) node[align=center, below] {$t_2$};
		\draw[ultra thick,decoration={snake,amplitude=0.06cm,segment length={\pathlen /7}},
		decorate] (0,0) arc (270:90:0.5cm);
		\draw [ ultra thick] (0,1) arc (90:-90:0.5cm);
		\end{tikzpicture} 
	\end{tabular}
\end{minipage}
\begin{minipage}[b]{0.55\textwidth}
\begin{equation}
\mathcal{I}(t_1,t_2)=\frac{3 \, \slashed{m}^2 \, \hbar \, H_0^3 \, \chi}{4\pi^2 } \int\frac{d\omega}{2\pi} \frac{e^{i\, \omega(t_2-t_1)}  }{\left(\slashed{m}^2+\omega^2\right)^2}
\end{equation}
\end{minipage}

\begin{minipage}[b]{0.5\textwidth}
\hspace{-1cm}
	\begin{tabular}{ cc } 
		\begin{tikzpicture}
		\filldraw[ultra thick] (-1.5,0)circle (2pt) node[align=center, below] {$t_1$} -- (-1,0);
		\filldraw[ultra thick] (-1,0) -- (0,0)circle (2pt);
		circle (2pt)
		\filldraw [  ultra thick  ] (0,0)circle (2pt) node[align=center, below]{$$} -- (1.5,0)circle (2pt) node[align=center, below] {$t_2$};
		\draw [ dashed, ultra thick] (0,0.5) circle [radius=0.5];
		\draw[->, ultra thick, dashed] (0,1)--(-0.01,1);
		\draw[ultra thick,decoration={snake,amplitude=0.06cm,segment length={\pathlen /7}},
		decorate] (0,0) arc (90:-90:0.5cm);
		\draw [ ultra thick] (0,-1) arc (270:90:0.5cm);
		\end{tikzpicture}
		&
		\begin{tikzpicture}
		\filldraw[ultra thick] (-1.5,0)circle (2pt) node[align=center, below] {$t_1$} -- (-1,0);
		\filldraw[ultra thick] (-1,0) -- (0,0)circle (2pt);
		circle (2pt)
		\filldraw [  ultra thick  ] (0,0)circle (2pt) node[align=center, below]{$$} -- (1.5,0)circle (2pt) node[align=center, below] {$t_2$};
		\draw[ultra thick,decoration={snake,amplitude=0.06cm,segment length={\pathlen /7}},decorate] (0,0) arc (90:-90:0.5cm);
		\draw [ ultra thick] (0,-1) arc (270:90:0.5cm);
		\draw[ultra thick,decoration={snake,amplitude=0.06cm,segment length={\pathlen /7}},
		decorate] (0,0) arc (270:90:0.5cm);
		\draw [ ultra thick] (0,1) arc (90:-90:0.5cm);
		\end{tikzpicture}   \\
	\end{tabular}
\end{minipage}
\begin{minipage}[b]{0.5\textwidth}

\begin{equation}
\mathcal{J}(t_1,t_2)=\frac{ \lambda \,\hbar \, H_0^5 \,\chi }{(8\pi^2)^2 }  \int\frac{d\omega}{2\pi} \frac{e^{i\, \omega(t_2-t_1)}  }{\left(\slashed{m}^2+\omega^2\right)^2}
\end{equation}
\end{minipage}

\vspace{2cm}
\begin{minipage}[b]{0.5\textwidth}
	\hspace{-1cm}
	\begin{tabular}{ cc } 
		\begin{tikzpicture}
		\filldraw[ultra thick] (-1.5,0)circle (2pt) node[align=center, below] {$t_1$} -- (-1,0);
		\filldraw[ultra thick] (-1,0) -- (0,0)circle (2pt) node[align=center, below]{$$};
		circle (2pt)
		\filldraw [  ultra thick  ] (1,0) -- (1.5,0)circle (2pt) node[align=center, below] {$t_2$};
		\draw [ ultra thick] (0,-1) arc (270:-270:0.5cm);
		\draw[ultra thick,decoration={snake,amplitude=0.06cm,segment length={\pathlen /7}},
		decorate] (0,0) arc (270:90:0.5cm);
		\draw [ ultra thick] (0,1) arc (90:-90:0.5cm);
		\draw[ultra thick,snake] (0,0) -- (1,0)circle ;
		\end{tikzpicture} 
		&
		\begin{tikzpicture}
		\filldraw[ultra thick] (-1.5,0)circle (2pt) node[align=center, below] {$t_1$} -- (-1,0);
		\draw[ultra thick,snake] (-1,0) -- (0,0)circle (2pt);
		\filldraw [  ultra thick  ] (0,0)circle (2pt) node[align=center, below]{$$} -- (1.5,0)circle (2pt) node[align=center, below] {$t_2$};
		\draw[ultra thick,decoration={snake,amplitude=0.06cm,segment length={\pathlen /7}},decorate] (0,0) arc (90:-90:0.5cm);
		\draw [ ultra thick] (0,-1) arc (270:90:0.5cm);
		\draw [ ultra thick] (0,1) arc (90:-270:0.5cm);
		\end{tikzpicture}      \\ 
		\begin{tikzpicture}
		\filldraw[ultra thick] (-1.5,0)circle (2pt) node[align=center, below] {$t_1$} -- (-1,0);
		\filldraw[ultra thick] (-1,0) -- (0,0)circle (2pt) node[align=center, below]{$$};
		\filldraw [  ultra thick  ] (1,0) -- (1.5,0)circle (2pt) node[align=center, below] {$t_2$};
		\draw [ ultra thick] (0,-1) arc (270:-270:0.5cm);
		\draw[ultra thick,snake] (0,0) -- (1,0)circle ;
		\draw [ dashed, ultra thick] (0,0.5) circle [radius=0.5];
		\draw[->, ultra thick, dashed] (0,1)--(-0.01,1);
		\end{tikzpicture}
		&
		\begin{tikzpicture}
		\filldraw[ultra thick] (-1.5,0)circle (2pt) node[align=center, below] {$t_1$} -- (-1,0);
		\draw[ultra thick,snake] (-1,0) -- (0,0)circle (2pt);
		\filldraw [  ultra thick  ] (0,0)circle (2pt) node[align=center, below]{$$} -- (1.5,0)circle (2pt) node[align=center, below] {$t_2$};
		\draw [ ultra thick] (0,-1) arc (270:-270:0.5cm);
		\filldraw[ultra thick] (0,0) -- (1.5,0)circle ;
		\draw [ dashed, ultra thick] (0,0.5) circle [radius=0.5];
		\draw[->, ultra thick, dashed] (0,1)--(-0.01,1);
		\end{tikzpicture}  	\\
	\end{tabular}
\end{minipage}
\begin{minipage}[b]{0.5\textwidth}
\begin{equation}
\mathcal{K}(t_1,t_2)= \frac{\lambda \, \hbar \, H_0^5\, \chi }{(8\pi^2)^2 } \int\frac{d\omega}{2\pi} \frac{e^{i\, \omega(t_2-t_1)}  }{\left(\slashed{m}^2+\omega^2\right)^2}
\end{equation}
\end{minipage}

\noindent where the symmetry factor of the last diagram have been taken as $\frac12$.
Lastly, there is a single diagram with two scalar loops: 

\vspace{1cm}
\begin{minipage}[b]{0.5\textwidth}
	\begin{tabular}{c} 
		\begin{tikzpicture}
		\filldraw[ultra thick] (-1.5,0)circle (2pt) node[align=center, below] {$t_1$} -- (-1,0);
		\draw[ultra thick,snake] (-1,0) -- (0,0)circle (2pt);
		\filldraw [  ultra thick  ] (0,0)circle (2pt) node[align=center, below]{};
		\draw [ultra thick] (0,-1) arc (270:-270:0.5cm);
		\filldraw[ultra thick] (1,0) -- (1.5,0)circle (2pt) node[align=center, below] {$t_2$} ;
		\draw [ultra thick] (0,0.5) circle [radius=0.5];
		\draw[ultra thick,snake] (0,0) -- (1,0);
		\end{tikzpicture} 	\\
	\end{tabular}
\end{minipage}
\begin{minipage}[b]{0.5\textwidth}
\begin{equation}
\mathcal{L}= \frac{ \lambda \, \hbar H_0^5\, \chi}{(8\pi^2)^2\,  4\slashed{m}^2} \int\frac{d\omega}{2\pi} \frac{e^{i\, \omega(t_2-t_1)}  }{\left(\slashed{m}^2+\omega^2\right)^2}
\end{equation}
\end{minipage}

where the symmetry factor of the last diagram have been taken as $\frac18$. Notice that now the ghost loops do not cancel closed $G$ loops due to non-matching numerical factors in the new vertices. We comment on this feature in the final section. It is important to note that the above comment does not apply to the bubble duagrams in the partition function; they stll cancel as described in \eqref{partition}. 
     
\section{Summary and conclusions}
In this work we elaborated on the path integral representation for the Langevin equation describing the infrared behaviour of a spectator scalar in de Sitter. Using it, we obtained simple Feynman Rules that allow for the straightforward computation of arbitrary unequal-time correlators of the field. Such quantities can also be computed via a perturbative expansion applied directly to the Langevin equation. As we demonstrated, the two approaches are equivalent but the former offers a more streamlined and efficient way to the final result, circumventing a lot of the the steps necessitated by the latter that involve a) breaking down the solution for $\phi$ in different orders, represented by the tree pre-graphs of section \ref{sec:direct-Langevin} and b) evaluating expectation values of their products by gluing the trees in all possible ways across their crossed tips. These steps become increasingly complex with the number of fields in the desired correlator and with perturbative order. On the contrary, the Feynman rules directly build any correlator out of two propagators and a small, fixed number of vertices. We expect that their utility will become even sharper when more than one field is involved. We also briefly considered backreaction by including the dependence of the noise amplitude on $\phi$, making it multiplicative noise, and calculated new contributions to the 2-point functions. In this case, the direct Langevin equation approach would have proven substantially more involved. 

Our Feynman Rules necessitated the introduction of ghost fields that contribute closed loops in the diagrams. In the case of additive noise $H=H_0$, ghost loops act to cancel closed $G$ loops. Such loops do not appear in the Langevin perturbative solution so ghosts are essential in ensuring that Feynman diagrams give the correct result. In the case of multiplicative noise $H=H(\phi)$, the contributions from ghost loops and closed $G$ loops do not add up to zero and hence contribute to the final result. This is a manifestation of the well known fact that when the noise is multiplicative, results depend on the discretization prescription of the Langevin equation which, in a continuum description, translates to the choice of the value of $\Theta(0)$ \cite{Lau-Lubensky}. Our formalism naturally picks the midpoint value $\Theta(0)=\frac{1}{2}$, corresponding to the Stratonovich prescription. Other prescriptions would also be possible to implement in a simple manner by adding appropriate ``spurious force'' terms to the potential. We leave an investigation of this point and an implementation of other prescriptions within our formalism for the future. 

For the time being, it is unclear to us which prescription would be the appropriate one if gravity is consistently included, with different prescriptions leading to different results for the correlators albeit suppressed by powers of $\chi \equiv \hbar G H_0^2/2\pi$. It is remarked in \cite{Vennin:2015hra} that this theoretical uncertainty should be commensurable to corrections to the leading stochastic picture. However, such corrections are now accessible and the uncertainty becomes relevant if one wants to go beyond the leading stochastic dynamics. As stressed in \cite{vanKampen}, the correct prescription for modelling dynamics via a stochastic differential equation can only be decided by either a first principles computation or other external physical considerations - mathematically all prescriptions ($\Theta(0)\in [0,1]$ in a continuum description) are equally admissible. In our case, the stochastic action entering the path integral must be a truncated version of the full underlying QFT action in the Schwinger-Keldysh formulation, see \cite{Moss:2016uix} for the spectator field case, and hence a particular prescription must be chosen from the underlying dynamics. Since no determinants appear in the QFT path integral, the Ito prescription seems favoured but this will need to be verified via a concrete computation. A similar reduction to that described in \cite{Moss:2016uix} including gravitational degrees of freedom, for which a QFT path integral has been derived \cite{Prokopec:2010be}, and its comparison to the stochastic $\Delta N$ formalism \cite{Vennin:2015hra, Enqvist:2008kt, Fujita:2013cna, Fujita:2014tja} would also be an interesting research pursuit for the future.        

\begin{center}
{\bf Acknowledgements}
\end{center}
GR wishes to thank Ian Moss, Tomislav Prokopec, Julien Serreau, Vincent Vennin and Sebastien Renaux-Petel for many interesting discussions on the stochastic description of inflationary fluctuations.

\end{document}